\documentclass[journal]{IEEEtran}
\ifCLASSINFOpdf
  \usepackage[pdftex]{graphicx}
\else
\fi
\usepackage{amsmath}
\usepackage{algorithmic}

\usepackage{array}

\ifCLASSOPTIONcompsoc
  \usepackage[caption=false,font=normalsize,labelfont=sf,textfont=sf]{subfig}
\usepackage[caption=false,font=footnotesize]{subfig}
\fi

\usepackage{stfloats}
\fnbelowfloat

\usepackage{amsmath,amsbsy,amssymb,amsthm}
\usepackage{mathrsfs}
\usepackage{graphicx}
\usepackage{longtable}
\usepackage{cite}
\usepackage{subcaption}
\usepackage{algorithm2e}
\usepackage{romannum}

\DeclareFontFamily{U}{calligra}{}
\DeclareFontShape{U}{calligra}{m}{n}{<->callig15}{}

\begin{document}

\title{Robust On-Line ADP-based Solution of a Class of Hierarchical Nonlinear Differential Game}
%
%

\author{Mohammad~Reza~Satouri, 
             Hamed~Kebriaei,~\IEEEmembership{Senior~Member,~IEEE,}
        Abolhassan~Razminia~
        and~Mohammad~Javad~Yazdanpanah
\thanks{M. R. Satouri and H. Kebriaei are with the School of Electrical and Computer Engineering, College of Engineering, University of Tehran, P.O. Box 4395-515, Tehran, Iran (e-mail: satouri.mr@ut.ac.ir ; kebriaei@ut.ac.ir ).

A. Razminia is with the Dynamical Systems \& Control (DSC) Research Lab.,
Department of Electrical Engineering, School of Engineering, Persian Gulf University, P.O. Box 75169, Boushehr, Iran (e-mail: razminia@pgu.ac.ir).

M. J. Yazdanpanah is with the School of Electrical and Computer Engineering, College of Engineering, University of Tehran, P.O. Box 4395-515, Tehran, Iran, and also with the Control \& Intelligent Processing Center of Excellence (e-mail: yazdan@ut.ac.ir).}}

%



\maketitle

\begin{abstract}
In this paper a hierarchical one-leader-multi-followers game for a class of continuous-time nonlinear systems with disturbance is investigated by a novel policy iteration reinforcement learning technique in which, the game model consists both of the zero-sum and nonzero-sum games, simultaneously. An adaptive dynamic programming (ADP) method is developed to achieve optimal control strategy under worst case of disturbance. This algorithm reduces the number of neural networks which are used for estimation for about thirty percent. Proposed algorithm uses neural networks to estimate value functions, control policies and disturbances. Convergence analysis of the estimations is investigated using Lyapunov theory and exploiting properties of the Nemytskii operator. Finally the simulation results will show effectiveness of developed ADP method. 
\end{abstract}


\newtheorem{theorem}{Theorem}
\newtheorem{definition}[theorem]{Definition}
\newtheorem{remark}[theorem]{Remark}
\newtheorem{lemma}[theorem]{Lemma}
%
\IEEEpeerreviewmaketitle

\section{Introduction}
%
%
%
%
\IEEEPARstart{N}{owadays}, the optimal control theory has penetrated in various disciplines from aerospace engineering to resource management \cite{a1,v1,v2,v3}. Based on its merits and importance, the scholars have developed different approaches to study the dynamic optimization problems \cite{v4}. One of the main trends in optimal control systems has been provided by Bellman, namely the optimality principle, and now extended to more complex versions. The adaptive dynamic programming (ADP) is one of these extensions whose essential root is the Hamilton-Jacobi-Bellman (HJB) equation \cite{v5,a3,a4,a5,a7,a8,a24,a25,a26,a27}.

Inspiring by the reinforcement learning (RL) techniques, the ADP resolves  some drawbacks of the classical dynamic programming (DP) like curse of dimensionality \cite{v6}. As a matter of fact, the kernel of the operation of ADP is  estimating the cost functional by a function approximating mechanism and then solving the DP problem forward in time. Based on its efficient performance, the ADP is widely used in numerous optimal control problems such as power system control \cite{a13,a14}, battery management \cite{a15,a16} and autopilot systems \cite{a17,a18}.

In a multi-agent system, when the control strategy of an agent affects the cost function of other players (e.g. through system's dynamic) one can model the problem as a dynamic game rather than a single optimal control problem. Usually, in a dynamic game with non-quadratic cost, nonlinear or unknown dynamics, the closed form of the solution which is known as the equilibrium point of the game, cannot be found and therefore, the ADP method can be utilized as a promising tool to estimate the solution.

Dynamic games can be classified based on the sum of the cost functions into zero-sum (ZS) and nonzero-sum (NZS) games \cite{a20,a21,a22,a23}.
In ZS games, two major approaches have been proposed in \cite{a34} where the underlying system is linear: the on-line algorithm based on the integral RL in which  some partial information of the dynamical system is needed, and the off-line algorithm based on the policy iteration (PI) wherein full information of the system is utilized. The extension of the approach for an unknown nonlinear dynamical system has been studied in \cite{a32}. In \cite{a33}, an ADP method has been proposed for input constraint ZS dynamic game. For the linear systems with partially measurable states in presence of various matched uncertainties, a robust ADP method for the ZS game has been explored in
\cite{a35}.
An on-line iterative ADP algorithm has been proposed in \cite{a36} with known disturbance matrix and unknown state and control matrices.

In contrast to the ZS games, where the goals of the players are fully opposed and the equilibrium point is known as the saddle point, in NZS games the equilibrium strategy of the game is defined as a strategy profile of the players in which, no player can lower its cost function by changing its control strategy, while the other players do not change their strategy. This equilibrium strategy is also known as Nash equilibrium point. In \cite{a39}, a PI-ADP approach for a NZS game with nonlinear discrete-time dynamic has been proposed and both on-line and off-line algorithms has been investigated. An on-line method for multi-player NZS games which requires complete knowledge of system dynamics has been presented in \cite{a20}. In \cite{a37}, an off-policy IRL method for a continuous-time multiplayer NZS game with unknown dynamics has been explored. Utilizing the actor-critic-identifier (ACI), without full information of the system dynamics is another important approach in NZS games. This technique is adopted in \cite{a38} to approximately solve a set of coupled HJB equations of the proposed $N$ player NZS game.

All of the mentioned research works which have used the ADP method to solve dynamic games, only considered the case that the players decide simultaneously. However, the problem becomes more complicated when the players decide in hierarchical scheme. In dynamic Nash games where the strategy of each player affects other players' costs and strategies, indirectly through the state of the system and the players have access to the state measurement, the Nash optimal strategy of the game can be obtained using ADP to solve HJB equations of the players, simultaneously.
 However, in a hierarchical (e.g. one-leader-multi-followers) game, since the strategies of the followers depend both on the strategy of the leader and the state of the system, finding the Stackelberg-Nash equilibrium point of the game and the proof of convergence of the ADP algorithm to the equilibrium point of the game become more complicated.

The basic type of hierarchical games is the leader-follower or Stackelberg game which was introduced by H. von Stackelberg in 1934 \cite{d1}. In this game, the leader decides first at the upper level (level 1), and in the lower level (level 2), the follower decides upon the decision of the leader is revealed \cite{a28,a29,a30,a31}. The leader-follower game has widespread applications in many engineering fields like smart grids and cognitive radio network \cite{b4,b5}.

Dynamic Stackelberg games are well studied in the case of linear dynamic and quadratic cost \cite{d2,d3,d4,b6}. However, studies on dynamic Stackelberg games with nonlinear dynamic and non quadratic cost are fairly scarce.
In \cite{d6} the authors proposed a heuristic iterative algorithm for solving dynamic Stackelberg game, however the convergence of the proposed algorithm was not proven.
In \cite{c1} discrete time dynamic Stackelberg games with general form of dynamic and cost is studied and dynamic programming approach is used to derive the Bellman equation and find the optimal solution. However, for continuous time systems the Bellman optimality principle leads to the well known HJB equation. In non-LQ continuous time problems like our case, the HJB equation cannot be solved analytically and ADP is a promising approach which can be utilized to obtain the value function and optimal strategy.

The ADP method has been utilized to solve non-LQ dynamic Nash-games in the literature like \cite{a32,a37}, nevertheless, to the best of authors' knowledge, the ADP has not been explored to solve nonlinear-non-quadratic Stackelberg games up to now. In this paper, an on-line ADP method is proposed to obtain the equilibrium point of a group of dynamic continuous time hierarchical nonlinear non-quadratic games. Further, the proposed algorithm is robust with respect to uncertainty which is considered as an exogenous disturbance. In the upper level (level 1) of the game, the leader acts first and plays a Stackelberg game with the followers. At the lower level (level 2), the followers play an $N$-player non-cooperative Nash game and decide simultaneously after the leader. All the players also play a ZS game with the disturbance.

Because of the non-simultaneous decision making in our proposed game, the strategy of the leader appears in
the Nash strategy of the followers, since the best response of each follower is a function of both the current state and the strategy of the leader. Such dependency imposes more difficulties when we want to calculate the equilibrium strategies of the players and also to prove the convergence of the ADP algorithm to the equilibrium point of the game. To overcome this issue, we have used Nemytskii operator and neural networks.
 In the proposed ADP algorithm, neural networks are used in actor-critic scheme. Update rules for critic and actor networks are obtained using gradient descent and adaptive control, respectively. Convergence analysis of the ADP method to equilibrium point of the game is performed using Lyapunov theory and exploiting the properties of Nemytskii operator.  In addition, the number of neural networks used in actor-critic method is reduced with respect to similar works \cite{a22,a35} which facilitates the training procedure and speeds up the convergence.

The contribution of this paper can be summarized as follows:
\begin{itemize}
\item
Using an ADP-based method to solve a class of differential hierarchical games with nonlinear dynamic and non-quadratic cost in presence of disturbance for the first time.
\item
Convergence analysis of ADP algorithm to the Stackelberg-Nash-Saddle equilibrium point of the proposed game model by guaranteeing the uniform ultimate boundedness (UUB) condition.
\item
Using Nemytskii operator in the ADP-based methods for the first time to overcome the complexity of computing the leader's optimal strategy.
\end{itemize}

The organization of the paper is as follows: In Section \Romannum{2} the problem is formulated and the proposed game model is discussed. Some preliminaries are reviewed in Section \Romannum{3}.  Section \Romannum{4} devotes to the main results in which the final solution is obtained in three serial steps. A numerical simulation in Section \Romannum{5}, verifies the obtained theoretical results. Concluding remarks are presented in Section \Romannum{6}.

%
%

\section{System Model and Problem Formulation}\label{sec2}

\begin{figure}[!ht]
\begin{center}
\includegraphics[scale=0.40]{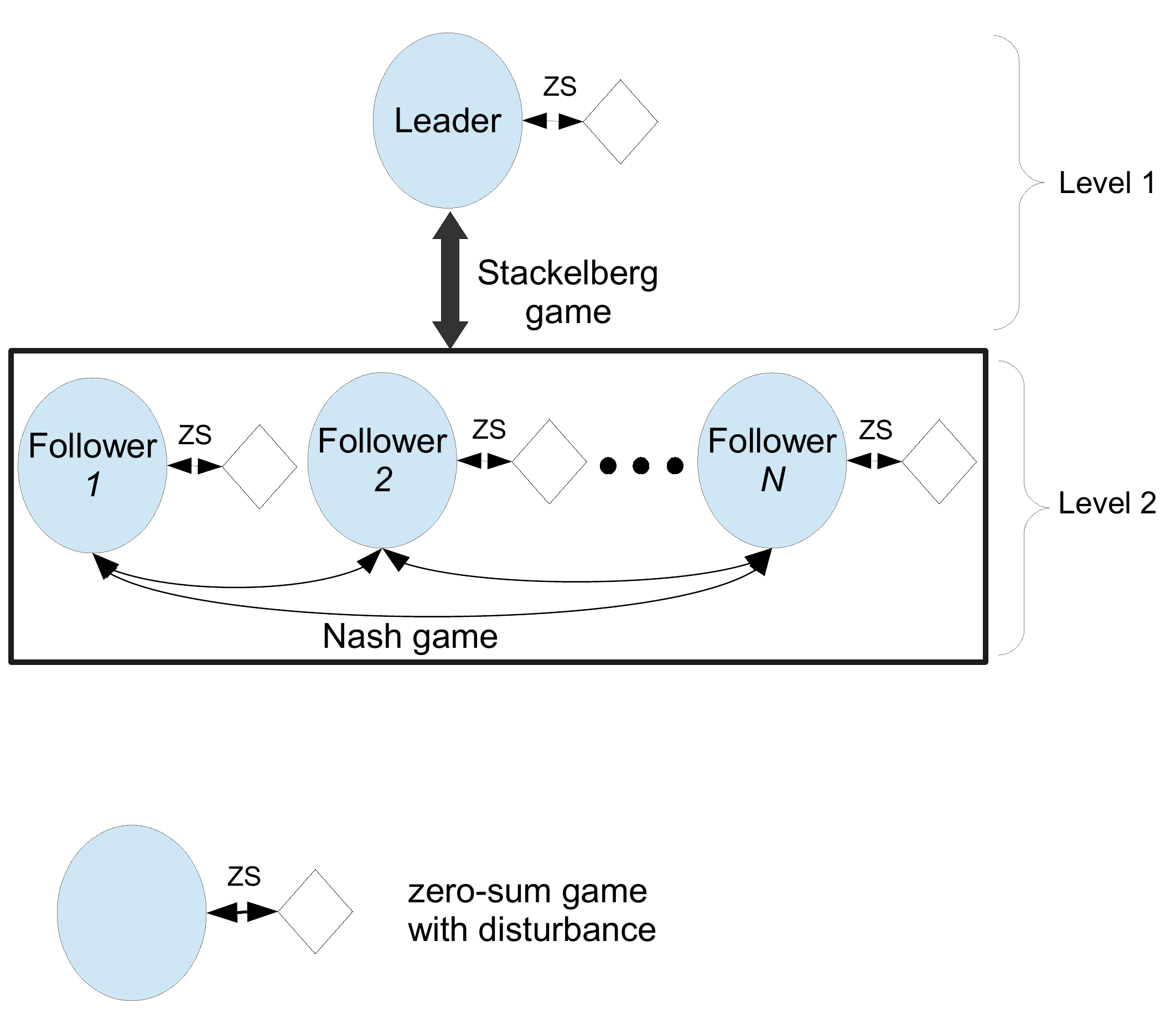}
\end{center}
\caption{Game model.}
\label{fig1}
\end{figure}

The general scheme and information structure of the proposed hierarchical game in this paper is adopted from \cite{b6} where the authors studied discrete time linear-quadratic hierarchical dynamic games. The game consists of $N+2$ players: one leader, $N$ followers and a disturbance. The players decide in a dynamic environment. The sate dynamic equation of the system is known to all players and also they have closed loop information about the state of the system. The strategy of all players appear in the state dynamic equation of the system and therefore, the objective function of all players are coupled indirectly through the state of the system. The disturbance is considered as a virtual player who wants to provide the worst case condition for the objective function of all players. In the lower level (level 2) of the proposed hierarchy, the followers play a dynamic Nash game with each other and also play a zero-sum game with the disturbance. The followers decide simultaneously and they don't have information about other followers' costs and strategies, while they know the leader's strategy. At this level, each follower independently updates its value function and  best response strategy  to the other followers' strategies and disturbance. In the higher level (level 1) of the game, the leader uses the obtained strategies of the followers as the reaction of the followers to play a ZS game with disturbance and find its min-max optimal strategy. Since there is a Stackelberg, a Nash and (N+1) ZS games in this model, the equilibrium point of the whole game is called Stackelberg-Nash-saddle equilibrium. A schematic diagram of the proposed game structure is shown in Fig. \ref{fig1}.
Consider a dynamical system with one leader and $N$ followers playing over a state space whose evolution dictated by  the following differential equation:
\begin{equation}\label{dynamic}
\dot{\mathbf{x}}(t)=\mathbf{f}(\mathbf{x})+\sum_{j=1}^N g_{j}(\mathbf{x})\mathbf{u}_{j} +p(\mathbf{x})\mathbf{\nu} +h(\mathbf{x})\boldsymbol{\boldsymbol{\omega}}
\end{equation}
where $ \mathbf{x}(t) \in \boldsymbol{\mathscr{X}} \subset \mathbb{R}^{n} $, $ \mathbf{u}_{j}(t) \in \boldsymbol{\mathscr{U}}_{j} \subset \mathbb{R}^{m_j} $, $ \boldsymbol{\nu}(t) \in \boldsymbol{\mathscr{N}} \subset \mathbb{R}^{\alpha} $ , and $ \boldsymbol{\omega}(t) \in \boldsymbol{\mathscr{W}} \subset \mathbb{R}^{w} $, are state vector, controls or actions of followers,  control or action of leader, and disturbance, respectively. Moreover, $\mathbf{f}: \mathbb{R}^{n}\rightarrow \mathbb{R}^n$ is a Lipschitz vector function with $\mathbf{f}(\mathbf{0})=\mathbf{0}$, and $g_j: \mathbb{R}^n\rightarrow \mathbb{R}^{n\times m_j}, j=1,...N$, $h: \mathbb{R}^n \rightarrow \mathbb{R}^{n\times \alpha}$, $p: \mathbb{R}^n\rightarrow \mathbb{R}^{n\times w}$ are some continuous matrices. In addition, similar to \cite{a19,a20,a38}, it is assumed that $ \Vert g_{i}(\mathbf{x})\Vert < b_{i} $ for $1\leq i\leq N $ , $ \Vert h(\mathbf{x}) \Vert <b_{h} $ and $ \quad \Vert p(\mathbf{x}) \Vert <b_{p} $. The leader's decision is based on followers' rationality. All players make decisions in such way to minimize their cost functionals in the worst case scenario generated by the disturbance with attenuation level $\gamma$ as follows:
\begin{align}
V_{i} &=\int_{t}^{\infty}\Bigg( Q_{i}(\mathbf{x}(\tau))+\sum_{j=1}^N \mathbf{u}_{j}^{T}(\tau)R_{ij}\mathbf{u}_{j}(\tau) \nonumber \\
&\qquad +\boldsymbol{\nu}^{T}(\tau)S_{i}\boldsymbol{\nu}(\tau) -\gamma^{2}\boldsymbol{\omega}^{T}(\tau)\boldsymbol{\omega}(\tau) \Bigg) \mathrm{d}\tau \nonumber \\
&= \int_{t}^{\infty}r_{i}(\mathbf{x}(\tau), \mathbf{u}_{1}(\tau), \mathbf{u}_{2}(\tau), \dots , \mathbf{u}_{N}(\tau), \boldsymbol{\nu}(\tau) , \boldsymbol{\omega}(\tau)) \mathrm{d}\tau \label{vi}
\end{align}
where $1\leq i\leq N$ are followers and $ i=N+1$ is the leader. Functions $ Q_{i}(\mathbf{x}(t))>0 $ are generally nonlinear and satisfy $ Q_{i}(\mathbf{x})\geq \vartheta_{i}\mathbf{x}^{T}\mathbf{x}$ for certain positive constants $\vartheta_{i}$, $ \gamma  $ is a scalar, and $ S_{i}>0, R_{ii}>0, R_{ij}\geq 0 $ are known and symmetric matrices. In the subsequent, the parameter $t$ is dropped for the sake of simplicity.

Based on the given points, the goal is to find a set of state feedback control policies which minimize cost functions in the worst case strategy of disturbance player. In the subsection 3.1 we solve the game in the lower level (level 2) for followers and then the obtained solution is used in subsection 3.2 to solve the leader's problem in the upper level (level 1) and complete the solution of the proposed bi-level game. Then in the subsection 3.3, the update rules for critic and actor networks are introduced and the convergence of the proposed algorithm is proven.

\section{Preliminary Definitions}\label{ss:mb}

\begin{definition}[ \cite{b1} Nash Equilibrium]
An $N$-tuple of strategies $\lbrace \boldsymbol{\gamma}^{1*},\boldsymbol{\gamma}^{2*},...,\boldsymbol{\gamma}^{N*}\rbrace $ with $ \boldsymbol{\gamma}^{i*}\in \Gamma^{i} $, where $\Gamma^{i} $ is $i$th player strategy set, $ i\in \mathbb{Z}^+$, is said to constitute a Nash equilibrium solution for $N$-person nonzero-sum finite game in extensive form, if the following $N$ inequalities are satisfied for all $ \boldsymbol{\gamma}^{i}\in \Gamma^{i} $, $ i\in \mathbb{Z}^+$ :
\begin{equation}
J^{i}(\boldsymbol{\gamma}^{1*},...,\boldsymbol{\gamma}^{N*})\leq J^{i}(\boldsymbol{\gamma}^{1*},...,\boldsymbol{\gamma}^{(i-1)*},\boldsymbol{\gamma}^{i},\boldsymbol{\gamma}^{(i+1)*},...,\boldsymbol{\gamma}^{N*})
\end{equation}
The $N$-tuple of quantities $\lbrace J^{1*},J^{2*},...,J^{N*}\rbrace$ is known as a Nash equilibrium outcome of the nonzero-sum finite game in extensive form.
\end{definition}

\begin{definition}[ \cite{b1} Stackelberg Game]
In a Stackelberg game a player (leader) has the ability to enforce his strategy on the other player (follower). In this game leader acts first and then follower reacts to leader's decision. A strategy $ \gamma^{1*} \in \Gamma^{1} $ is called a Stackelberg equilibrium strategy for the leader, if
\begin{align}
J^{1*}&=\max_{\gamma^{2}\in R^{2}(\gamma^{1*})} J^{1}(\gamma^{1*}, \gamma^{2})\nonumber \\
&=\min_{\gamma^{1} \in \Gamma^{1}} \max_{\gamma^{2} \in R^{2}(\gamma^{1})}J^{1}(\gamma^{1}, \gamma^{2})
\end{align}
where $ \Gamma^{1} $ , $ J^{1}(\gamma^{1}, \gamma^{2}) $ and $ \Gamma^{2} $ , $ J^{2}(\gamma^{1}, \gamma^{2}) $ are pure strategy spaces and costs of leader and follower, respectively. Also the set $ R^{2}(\gamma^{1}) \subset \Gamma^{2} $ defined for each $ \gamma^{1} \in \Gamma^{1} $ by
\begin{equation}
R^{2}(\gamma^{1})=\{ \xi \in \Gamma^{2} : J^{2}(\gamma^{1}, \xi)\leq J^{2}(\gamma^{1}, \gamma^{2}), \; \forall \gamma^{2}\in \Gamma^{2} \}
\end{equation}
is the optimal response set of follower to the strategy $ \gamma^{1} \in \Gamma^{1} $ of leader.
\end{definition}
\begin{definition}[ \cite{b1} Differential Game]
Consider a game with two players who decide in a dynamic environment as follows
\begin{equation}
\dot{x}(t)=f(t, x(t), u^{1}(t), u^{2}(t))
\end{equation}
and the cost functionals of the players $ \forall t \in [t_0,T]$ defined as:
\begin{equation}
L^{i}(u^{1}, u^{2})=\int_{t_0}^{T}g^{i}(t, x(t), u^{1}(t), u^{2}(t))\mathrm{d}t+q^{i}(x(T))
\end{equation}
 where $[t_0, T]$ denotes the prescribed duration of the game. If the players decide simultaneously, then the game is called differential Nash game or simply differential game and if they decide in leader-follower scheme, the game is called differential Stackelberg game. 
\end{definition}

\begin{definition}[ \cite{b2} Nemytskii Operator]
Consider a set $\Psi$ which is a measure or metric space and let $ X $ and $Y$  be two Hausdorff topological spaces. The Nemytskii operator is defined by:
\begin{equation}
N_{f}(\mathbf{u})(\mathbf{z})\triangleq f(\mathbf{z},\mathbf{u}(\mathbf{z})) \quad \forall \mathbf{z} \in \Psi 
\end{equation}
in which  $ \mathbf{f}:\Psi \times X\rightarrow Y $.
\end{definition}

As can be seen from the definition, the Nemytskii operator assigns the $Y$-valued $ \mathbf{z} \longmapsto f(\mathbf{z},\mathbf{u}(\mathbf{z}))$ to each function $ \mathbf{u}: \Psi \rightarrow X$. Another concept is the uniformly ultimately boundedness (UUB) property of a signal.

\begin{definition}[ \cite{b3} UUB]
The solutions of the nonlinear dynamical system $ \dot{\mathbf{x}}(t)=\mathbf{f}(t,\mathbf{x})$ with $\mathbf{x}(t_{0})=\mathbf{x}_{0}$, are uniformly ultimately bounded with ultimate bound $b$ if there exists a positive constant $c$ , independent of $t_{0}$, and for every $ a\in (0,c)$, there is a $ T(a,b)\geq 0$, such that:
\begin{equation}
\Vert \mathbf{x}(t_{0})\Vert \leq a \quad \Rightarrow \quad \Vert \mathbf{x}(t)\Vert \leq b,\quad  \forall t\geq t_{0}+T .
\end{equation}
\end{definition}

\section{Main Results}\label{main}

According to cost functionals \eqref{vi}, the optimal value functions for all the players in presence of disturbance are obtained using min-max optimization as follows:
\begin{align}
V_{i}^{\star}(\mathbf{x}, \mathbf{u}_{\bullet}, \boldsymbol{\nu}, \boldsymbol{\omega}_{i}) =\min_{\mathbf{u}_{i}}\max_{\boldsymbol{\omega}_{i}}\int_{t}^{\infty}\Bigg( Q_{i}(\mathbf{x}) \nonumber \\
+\sum_{j=1}^{N} \mathbf{u}_{j}^{T}R_{ij}\mathbf{u}_{j}+\boldsymbol{\nu}^{T}S_{i}\boldsymbol{\nu} -\gamma^{2}\boldsymbol{\omega}_{i}^{T}\boldsymbol{\omega}_{i} \Bigg) \mathrm{d}\tau \label{vio}
\end{align}
\begin{align}
&V_{N+1}^{\star}(\mathbf{x}, \mathbf{u}_{\bullet}, \boldsymbol{\nu} , \boldsymbol{\omega}_{N+1}) =\min_{\boldsymbol{\nu}}\max_{\boldsymbol{\omega}_{N+1}}\int_{t}^{\infty}\Bigg( Q_{N+1}(\mathbf{x})\nonumber \\
&\quad +\sum_{j=1}^{N} \mathbf{u}_{j}^{T}R_{(N+1)j}\mathbf{u}_{j}+\boldsymbol{\nu}^{T}S_{N+1}\boldsymbol{\nu} -\gamma^{2}\boldsymbol{\omega}_{N+1}^{T}\boldsymbol{\omega}_{N+1} \Bigg) \mathrm{d}\tau \label{vno}
\end{align}
where $ \mathbf{u}_{\bullet}=\lbrace \mathbf{u}_{1}, \mathbf{u}_{2}, \dots , \mathbf{u}_{N}\rbrace $ and $1\leq i\leq N$.
For the optimal set $\mathbf{u}_{\bullet}^*,\boldsymbol{\nu}^*,\boldsymbol{\omega}^*$ the  HJB equation holds as follows:
\begin{equation}
\mathcal{H}_{i}(\mathbf{x},\nabla_{\mathbf{x}} V_{i}^*, \mathbf{u}_{\bullet}^*, \boldsymbol{\nu}^* , \boldsymbol{\omega}^* ) +\frac{\partial V_{i}}{\partial t} =0
\end{equation}
where
\begin{align}
&\mathcal{H}_{i}(\mathbf{x},\nabla_{\mathbf{x}} V_{i}, \mathbf{u}_{\bullet}, \boldsymbol{\nu} , \boldsymbol{\omega} )=r_{i}(\mathbf{x}, \mathbf{u}_{\bullet}, \boldsymbol{\nu} , \boldsymbol{\omega} )+(\nabla_{\mathbf{x}} V_{i})^{T}\Bigg( \mathbf{f}(\mathbf{x}) \nonumber \\
&\qquad +\sum_{j=1}^N g_{j}(\mathbf{x})\mathbf{u}_{j} +h(\mathbf{x})\boldsymbol{\omega}+p(\mathbf{x})\boldsymbol{\nu} \Bigg)\label{hamiltonian}
\end{align}

In proceeding the aim is to find the optimal policies of followers and leader and then solving the whole game. Also from now on, the subscript $\mathbf{x}$ in $\nabla_{\mathbf{x}}$ is dropped for the sake of simplicity.

\subsection{Level 2: Followers-Disturbance Game}\label{subfollowers}


According to \eqref{vi} the cost functionals are convex and concave with respect to controls and disturbance, respectively. The minimum and maximum of the Hamiltonian can be directly computed by the stationary condition as follows:
\begin{align}
\frac{\partial \mathcal{H}_{i}}{\partial \mathbf{u}_{i}} &=0 \quad \Rightarrow \quad \mathbf{u}_{i}^{\star}(\mathbf{x})=-\frac{1}{2}R_{ii}^{-1}g_{i}^{T}(\mathbf{x})\nabla V_{i} \label{st-u} \\
\frac{\partial \mathcal{H}_{i}}{\partial \boldsymbol{\omega}_{i}} &=0 \quad \Rightarrow \quad \boldsymbol{\omega}_{i}^{\star}(\mathbf{x})=\frac{1}{2}\gamma^{-2}h^{T}(\mathbf{x})\nabla V_{i} \label{st-w}
\end{align}
Substituting \eqref{st-u} and \eqref{st-w} into \eqref{hamiltonian}, $N$ coupled Hamilton-Jacobi equations can be obtained:
\begin{align}
& Q_{i}(\mathbf{x})+\frac{1}{4}\sum_{j=1}^{N} (\nabla V_{j})^{T}g_{j}(\mathbf{x})R_{jj}^{-1}R_{ij}R_{jj}^{-1}g_{j}^{T}(\mathbf{x})\nabla V_{j}\nonumber \\
& -\frac{\gamma^{-2}}{4}(\nabla V_{i})^{T}h(\mathbf{x})h^{T}(\mathbf{x})(\nabla V_{i})+ \boldsymbol{\nu}^{T}S_{i}\boldsymbol{\nu}\nonumber \\
& +(\nabla V_{i})^{T}\left( \mathbf{f}(\mathbf{x})+\sum_{j=1}^{N} g_{j}(\mathbf{x})\mathbf{u}_{j}+p(\mathbf{x})\boldsymbol{\nu} +h(\boldsymbol{x})\boldsymbol{\omega} \right)=0 \label{ephji}
\end{align}
Since neither $ V_{i}$'s nor optimal control strategies can be computed in a closed form, a conventional way is to approximate the value functions and control strategies with general function approximators (GFAs). A common tool that is employed in implementing the GFA, is neural networks.
Therefore, the solutions of \eqref{ephji} on a compact set $ \Omega \subseteq \mathbb{R}^{n} $ which contains the origin are approximated by NNs as follows:
\begin{equation}\label{appvi}
V_{i}(\mathbf{x},\boldsymbol{\nu})=\mathbf{W}_{v_{i}}^{T}\boldsymbol{\phi}_{v_{i}}(\mathbf{x},\boldsymbol{\nu})+\epsilon_{v_{i}}(\mathbf{x},\boldsymbol{\nu}), \quad 1\leq i\leq N,
\end{equation}
where $ \boldsymbol{\phi}_{v_{i}}(.,.)\in \mathbb{R}^{\kappa} $ and $ \epsilon_{v_{i}} $ are NN activation functions and NN approximation errors, respectively. In \cite{a19} it has been shown that always there exists a number of hidden-layer neurons which makes the controls admissible and $ \epsilon_{v_{i}}$'s tend to zero as the number of hidden-layer neurons tends to infinity. It should be mentioned that because of hierarchical nature of the game,  $ V_{i} $s , $ 1\leq i\leq N $ are obtained as functions of $\mathbf{x}$ and $\boldsymbol{\nu} $. After calculating $ \boldsymbol{\nu}$ for the leader, the value functions would obtain in terms of $\mathbf{x}$ and then, it is be substituted in control policies of the followers.

In proceeding the new variables:
\begin{align}
C_{j}^{i}&=\nabla \boldsymbol{\phi}_{v_{j}}g_{j}(\mathbf{x})R_{jj}^{-1}R_{ij}R_{jj}^{-1}g_{j}^{T}(\mathbf{x})\nabla \boldsymbol{\phi}_{v_{j}}^{T}\nonumber \\
E_{i}&=\gamma^{-2}\nabla \boldsymbol{\phi}_{v_{i}}h(\mathbf{x})h^{T}(\mathbf{x})\nabla \boldsymbol{\phi}_{v_{i}}^{T}\nonumber \\
D_{i}^{j}&=\nabla \boldsymbol{\phi}_{v_i}g_{j}(\mathbf{x})R_{jj}^{-1}g_{j}^{T}(\mathbf{x})\nabla \boldsymbol{\phi}_{v_j}^{T}\nonumber
\end{align}
are introduced for simplicity. Substituting the value function approximations from \eqref{appvi} into \eqref{ephji} yields:
\begin{align}
& \frac{1}{4}\sum_{j=1}^{N} \mathbf{W}_{v_{j}}^{T}C_{j}^{i}\mathbf{W}_{v_{j}}-\frac{1}{4}\mathbf{W}_{v_{i}}^{T}E_{i}\mathbf{W}_{v_{i}}+Q_{i}(\mathbf{x})+\boldsymbol{\nu}^{T}S_{i}\boldsymbol{\nu} \nonumber \\
&\quad +\mathbf{W}_{v_{i}}^{T}\Bigg( \nabla \boldsymbol{\phi}_{v_{i}}\mathbf{f}(\mathbf{x})-\frac{1}{2}\sum_{j=1}^{N} D_{i}^{j}\mathbf{W}_{v_{j}} +\nabla \boldsymbol{\phi}_{v_{i}}p(\mathbf{x})\boldsymbol{\nu}\nonumber \\
&\quad +\frac{1}{2} E_{i}\mathbf{W}_{v_{i}} \Bigg)=\epsilon_{HJ_{i}}(\mathbf{x},\boldsymbol{\nu})\label{ephji_main}
\end{align}
where $\epsilon_{HJ_{i}}(\mathbf{x},\boldsymbol{\nu})$ is the error incurred by replacing the actual value functions with the approximated ones \eqref{appvi}. Now, according to \eqref{appvi}, the equations \eqref{st-u} and \eqref{st-w} can be rewritten as:
\begin{align}
\mathbf{u}_{i}(\mathbf{x},\boldsymbol{\nu})&=-\frac{1}{2}R_{ii}^{-1}g_{i}^{T}(\mathbf{x})\nabla \boldsymbol{\phi}_{v_{i}}^{T}(\mathbf{x},\boldsymbol{\nu})\mathbf{W}_{v_{i}} , \quad 1\leq i\leq N \\
\boldsymbol{\omega}_{i}(\mathbf{x},\boldsymbol{\nu})&=\frac{1}{2}\gamma^{-2}h^{T}(\mathbf{x})\nabla \boldsymbol{\phi}_{v_{i}}^{T}(\mathbf{x},\boldsymbol{\nu})\mathbf{W}_{v_{i}} , \quad 1\leq i\leq N
\end{align}
Since the ideal weights $ \mathbf{W}_{v_i} $'s are unknown, $ V_{i} $s and the control policies and disturbances should be defined in the form of critic and action neural networks, respectively as follows for $ 1\leq i\leq N $:
\begin{align}
\hat{V}_{i}(\mathbf{x},\boldsymbol{\nu})&=\hat{\mathbf{W}}_{v_{i}}^{T}\boldsymbol{\phi}_{v_{i}}(\mathbf{x},\boldsymbol{\nu}), \label{critic NN}\\
\hat{\mathbf{u}}_{i}(\mathbf{x},\boldsymbol{\nu})&=-\frac{1}{2}R_{ii}^{-1}g_{i}^{T}(\mathbf{x})\nabla \boldsymbol{\phi}_{v_{i}}^{T}(\mathbf{x},\boldsymbol{\nu})\hat{\mathbf{W}}_{a_{i}}, \label{u_hat} \\
\hat{\boldsymbol{\omega}}_{i}(\mathbf{x},\boldsymbol{\nu})&=\frac{1}{2}\gamma^{-2}h^{T}(\mathbf{x})\nabla \boldsymbol{\phi}_{v_{i}}^{T}(\mathbf{x},\boldsymbol{\nu})\hat{\mathbf{W}}_{a_{i}}\label{w_hat}
\end{align}
Substituting \eqref{u_hat} and \eqref{w_hat} in \eqref{ephji}, the following expression is obtained:
\begin{align}
& \frac{1}{4}\sum_{j=1}^{N}\hat{\mathbf{W}}_{a_{j}}^{T} C_{j}^{i}\hat{\mathbf{W}}_{a_{j}}-\frac{1}{4}\hat{\mathbf{W}}_{a_{i}}^{T} E_{i}\hat{\mathbf{W}}_{a_{i}}+Q_{i}(\mathbf{x})+\hat{\boldsymbol{\nu}}^{T}S_{i}\hat{\boldsymbol{\nu}} \nonumber \\
&\quad +\hat{\mathbf{W}}_{v_{i}}^{T} \Bigg( \nabla \boldsymbol{\phi}_{v_{i}}\mathbf{f}(\mathbf{x})-\sum_{j=1}^{N}\frac{1}{2}D_{i}^{j}\hat{\mathbf{W}}_{a_{j}}+\nabla \boldsymbol{\phi}_{v_{i}}p(\mathbf{x})\hat{\boldsymbol{\nu}}\nonumber \\
&\quad +\frac{1}{2}E_{i}\hat{\mathbf{W}}_{a_{i}} \Bigg)=e_{i}(\mathbf{x},\boldsymbol{\nu}) \label{ei}
\end{align}
where $ e_{i}(\mathbf{x},\boldsymbol{\nu}) $ is the error incurred by replacing ideal weights of NNs with their approximate values.

\subsection{Level 1: Leader-Disturbance Game}\label{subleader}

Similar to the previous subsection, the value function and critic NN of the leader can be approximated as follows:
\begin{equation}
\begin{cases}
V_{N+1}(\mathbf{x})&=\mathbf{W}_{v_{N+1}}^{T}\boldsymbol{\phi}_{v_{N+1}}+\epsilon_{v_{N+1}}(\mathbf{x})  \\
\hat{V}_{N+1}(\mathbf{x})&=\hat{\mathbf{W}}_{v_{N+1}}^{T}\boldsymbol{\phi}_{v_{N+1}}(\mathbf{x}) \label{critic NN_n+1}
\end{cases}
\end{equation}
After applying stationary condition of $ \boldsymbol{\omega} $ to the leader's Hamiltonian function similar to \eqref{w_hat}, one can get:
\begin{equation}
\hat{\boldsymbol{\omega}}_{N+1}(\mathbf{x})=\frac{1}{2}\gamma^{-2}h^{T}(\mathbf{x})\nabla \boldsymbol{\phi}_{v_{N+1}}^{T}(\mathbf{x})\hat{\mathbf{W}}_{a_{N+1}}  \label{w_hat_n+1}
\end{equation}
Substituting \eqref{critic NN_n+1} and \eqref{w_hat_n+1} into the leader's Hamiltonian, one can get:
\begin{align}
& \frac{1}{4}\sum_{j=1}^{N} \mathbf{W}_{v_{j}}^{T}C_{j}^{N+1}\mathbf{W}_{v_{j}}-\frac{1}{4}\mathbf{W}_{v_{N+1}}^{T}E_{N+1}\mathbf{W}_{v_{N+1}}\nonumber \\
&+\check{\boldsymbol{\nu}}^{T}S_{N+1}\check{\boldsymbol{\nu}}+Q_{N+1}(\mathbf{x})+\mathbf{W}_{v_{N+1}}^{T}\Bigg(\nabla \boldsymbol{\phi}_{v_{N+1}}\mathbf{f}(\mathbf{x})\nonumber \\
&-\frac{1}{2}\sum_{j=1}^{N} D_{N+1}^{j}\mathbf{W}_{v_{j}}+\nabla \boldsymbol{\phi}_{v_{N+1}}p(\mathbf{x})\check{\boldsymbol{\nu}}\nonumber \\
&+\frac{1}{2}E_{N+1}\mathbf{W}_{v_{N+1}}\Bigg) =\epsilon_{HJ_{N+1}}(\mathbf{x},\check{\boldsymbol{\nu}}) \label{ephjN+1}
\end{align}
and
\begin{align}
& \frac{1}{4}\sum_{j=1}^{N} \hat{\mathbf{W}}_{a_{j}}^{T}C_{j}^{N+1}\hat{\mathbf{W}}_{a_{j}} -\frac{1}{4}\hat{\mathbf{W}}_{a_{N+1}}^{T}E_{N+1}\hat{\mathbf{W}}_{a_{N+1}}\nonumber \\
&+ \hat{\boldsymbol{\nu}}^{T}S_{N+1}\hat{\boldsymbol{\nu}}+Q_{N+1}(\mathbf{x})+\hat{\mathbf{W}}_{v_{N+1}}^{T} \Bigg(\nabla \boldsymbol{\phi}_{v_{N+1}}\mathbf{f}(\mathbf{x})\nonumber \\
&-\frac{1}{2}\sum_{j=1}^{N} D_{N+1}^{j}\hat{\mathbf{W}}_{a_{j}}+\nabla \boldsymbol{\phi}_{v_{N+1}}p(\mathbf{x})\hat{\boldsymbol{\nu}}\nonumber \\
&+\frac{1}{2}E_{N+1}\hat{\mathbf{W}}_{a_{N+1}}\Bigg)=e_{N+1}(\mathbf{x},\hat{\boldsymbol{\nu}}) \label{en+1}
\end{align}
where $ \check{\boldsymbol{\nu}}$ and $ \hat{\boldsymbol{\nu}}$ are functions of exact and approximated NNs weights, respectively. In proceeding new variables
\begin{align}
B_{j}^{N+1}&=g_{j}(\mathbf{x})R_{jj}^{-1}R_{(N+1)j}R_{jj}^{-1}g_{j}^{T}(\mathbf{x})\nonumber \\
\check{\Pi}&=p^{T}(\mathbf{x})\nabla \boldsymbol{\phi}_{v_{N+1}}^{T}\mathbf{W}_{v_{N+1}}\nonumber \\
\hat{\Pi}&=p^{T}(\mathbf{x})\nabla \boldsymbol{\phi}_{v_{N+1}}^{T}\hat{\mathbf{W}}_{v_{N+1}}\nonumber \\
\check{A}_{j}^{N+1}&=\mathbf{W}_{v_j}\mathbf{W}_{v_{N+1}}^{T}\nabla \boldsymbol{\phi}_{v_{N+1}}g_{j}(\mathbf{x})R_{jj}^{-1}g_{j}^{T}(\mathbf{x})\nonumber \\
\hat{A}_{j}^{N+1}&=\hat{\mathbf{W}}_{a_j}\hat{\mathbf{W}}_{v_{N+1}}^{T}\nabla \boldsymbol{\phi}_{v_{N+1}} g_{j}(\mathbf{x})R_{jj}^{-1}g_{j}^{T}(\mathbf{x})\nonumber
\end{align}
are introduced for simplicity.
Now stationary condition of $ \check{\boldsymbol{\nu}}$ is applied to \eqref{ephjN+1}:
\begin{align}
\frac{\partial \mathcal{H}_{N+1}}{\partial \check{\boldsymbol{\nu}}}&=\check{\Pi}+\Bigg[\frac{1}{2}\sum_{j=1}^{N}\left( \frac{\partial \nabla \boldsymbol{\phi}_{v_{j}}(\mathbf{x},\check{\boldsymbol{\nu}})}{\partial \check{\boldsymbol{\nu}}}\right)^{T}\mathbf{W}_{v_{j}}\mathbf{W}_{v_{j}}^{T}\nonumber \\
&\times \nabla \boldsymbol{\phi}_{v_{j}}(\mathbf{x},\check{\boldsymbol{\nu}})B_{j}^{N+1}-\check{A}_{j}^{N+1}\Bigg] +2S_{N+1}\check{\boldsymbol{\nu}}=0
\end{align}
Thus
\begin{align}\label{nucheck}
\check{\boldsymbol{\nu}}&=-\frac{1}{2} S_{N+1}^{-1}\Bigg[ -\frac{1}{2}\sum_{j=1}^{N}\left( \frac{\partial \nabla \boldsymbol{\phi}_{v_{j}}(\mathbf{x},\check{\boldsymbol{\nu}})}{\partial \check{\boldsymbol{\nu}}}\right)^{T}\Bigg(\check{A}_{j}^{N+1}\nonumber \\
&\quad -\mathbf{W}_{v_{j}}\mathbf{W}_{v_{j}}^{T}\nabla \boldsymbol{\phi}_{v_{j}}(\mathbf{x},\check{\boldsymbol{\nu}})B_{j}^{N+1}\Bigg)+\check{\Pi} \Bigg]  
\end{align}
similar to \eqref{nucheck}:
\begin{align}\label{nuhat}
\hat{\boldsymbol{\nu}}&=-\frac{1}{2} S_{N+1}^{-1}\Bigg[ -\frac{1}{2}\sum_{j=1}^{N}\left( \frac{\partial \nabla \boldsymbol{\phi}_{v_{j}}(\mathbf{x},\hat{\boldsymbol{\nu}})}{\partial \hat{\boldsymbol{\nu}}}\right)^{T}\Bigg(\hat{A}_{j}^{N+1}\nonumber \\
&\quad -\hat{\mathbf{W}}_{a_{j}}\hat{\mathbf{W}}_{a_{j}}^{T}\nabla \boldsymbol{\phi}_{v_{j}}(\mathbf{x},\hat{\boldsymbol{\nu}})B_{j}^{N+1}\Bigg)+\hat{\Pi} \Bigg]  
\end{align}
To obtain the optimal strategy of the leader, we need to solve \eqref{nuhat} for $ \hat{\boldsymbol{\nu}} $. However, \eqref{nuhat} cannot be solved analytically with respect to $ \hat{\boldsymbol{\nu}} $. Therefore, to obtain the solution iteratively, we rewrite \eqref{nuhat} as an iterative equation in which, the
fixed point of the iterative equation is the optimal strategy of the leader:
\begin{align}\label{nuchecki}
\hat{\boldsymbol{\nu}}_{k+1}&=-\frac{1}{2} S_{N+1}^{-1}\Bigg[ -\frac{1}{2}\sum_{j=1}^{N}\left( \frac{\partial \nabla \boldsymbol{\phi}_{v_{j}}(\mathbf{x},\hat{\boldsymbol{\nu}}_{k})}{\partial \hat{\boldsymbol{\nu}}_{k}}\right)^{T}\Bigg(\hat{A}_{j}^{N+1}\nonumber \\
&\quad - \mathbf{W}_{v_{j}}\mathbf{W}_{v_{j}}^{T}\nabla \boldsymbol{\phi}_{v_{j}}(\mathbf{x},\hat{\boldsymbol{\nu}}_{k})B_{j}^{N+1}\Bigg)+\hat{\Pi} \Bigg]
\end{align}
To prove existence and uniqueness of the fixed point of iterative equation \eqref{nuchecki}, the Nemytskii operator is utilized and\eqref{nuhat} is rewritten as follows:
\begin{equation}\label{Nemop-nucheck}
\hat{\boldsymbol{\nu}}(\mathbf{x})=\Xi(\mathbf{x},\hat{\boldsymbol{\nu}}(\mathbf{x}))+\xi(\mathbf{x})
\end{equation}
where
\begin{equation*}
\xi(\mathbf{x})=-\frac{1}{2} S_{N+1}^{-1}p^{T}(\mathbf{x})\nabla \boldsymbol{\phi}_{v_{N+1}}^{T}\hat{\mathbf{W}}_{v_{N+1}},
\end{equation*}
where $ \xi(\mathbf{x})\in W^{1,\beta}({\Omega}) $, $1\leq \beta \leq \infty$ and $W^{1,\beta}({\Omega})$ is a Sobolev space and $ \Xi: \Omega \times \mathbb{R^{\alpha}}\rightarrow \mathbb{R^{\alpha}} $ is a Caratheodory function such that the Nemytskii operator $N_{\Xi}$ is continuous and bounded \cite{b2}. In the proceeding lemma the condition in which the whole game has a unique Stackelberg-Nash-saddle equilibrium is introduced.
\begin{lemma}\label{lem}
Equation \eqref{Nemop-nucheck} has a unique solution $ \hat{\boldsymbol{\nu}} \in W^{1,\beta}({\Omega})$ if Lipschitz coefficient $\Vert \Xi \Vert_{Lip}$ is less than one.
\end{lemma}
\begin{IEEEproof}
Suppose that there exist solutions $\hat{\boldsymbol{\nu}}_{1}$ and $ \hat{\boldsymbol{\nu}}_{2}$, so from \eqref{Nemop-nucheck} it is concluded that $ \Xi(\mathbf{x},\check{\boldsymbol{\nu}}_1)-\Xi(\mathbf{x},\hat{\boldsymbol{\nu}}_2)=\hat{\boldsymbol{\nu}}_{1}-\hat{\boldsymbol{\nu}}_{2}$. Thus
\begin{equation}
\Xi(\mathbf{x},\hat{\boldsymbol{\nu}}_1)-\Xi(\mathbf{x},\hat{\boldsymbol{\nu}}_2)=\hat{\boldsymbol{\nu}}_{1}-\hat{\boldsymbol{\nu}}_{2}\leq \Vert \Xi \Vert_{Lip}\big( \hat{\boldsymbol{\nu}}_{1}-\hat{\boldsymbol{\nu}}_{2}\big)
\end{equation}
Now if the Lipschitz coefficient $\Vert \Xi \Vert_{Lip}$ is less than one, the assumption of existing more than one solution is incorrect.
\end{IEEEproof}
According to the condition in Lemma \ref{lem}, the Nemytskii operator $N_{\Xi}$ has a unique fixed point and then  $ \hat{\boldsymbol{\nu}} $ can be calculated iteratively. The following NN is used for approximating $\hat{\boldsymbol{\nu}}$:\\
\begin{equation}\label{nuhatii}
\hat{\boldsymbol{\nu}}=\hat{W}_{\nu}^{T}\boldsymbol{\phi}_{\nu}
\end{equation}
where $\hat{W}_{\nu} \in \mathbb{R}^{\mu \times \alpha}$ and $\boldsymbol{\phi}_{\nu} \in \mathbb{R}^{\mu}$. Then the gradient descent is used for updating $\hat{W}_{\nu}$:
\begin{align}
&\hat{W}_{\nu}^{T}\boldsymbol{\phi}_{\nu}+\frac{1}{2} S_{N+1}^{-1}\Bigg[ -\frac{1}{2}\sum_{j=1}^{N}\left( \frac{\partial \nabla \boldsymbol{\phi}_{v_{j}}(\mathbf{x},\hat{\boldsymbol{\nu}})}{\partial \hat{\boldsymbol{\nu}}}\right)^{T}\Bigg(\hat{A}_{j}^{N+1}\nonumber \\
&-\hat{\mathbf{W}}_{a_{j}}\hat{\mathbf{W}}_{a_{j}}^{T}\nabla \boldsymbol{\phi}_{v_{j}}(\mathbf{x},\hat{\boldsymbol{\nu}})B_{j}^{N+1}\Bigg)+\hat{\Pi} \Bigg]=e_{\nu} \label{nuhati}
\end{align}
and
\begin{equation*}
\dot{\hat{W}}_{\nu}=-\varrho \frac{\partial e_{\nu}^{T}e_{\nu}}{\partial \hat{W}_{\nu}}
\end{equation*}
Where $ \varrho $ is the learning rate. The proof of convergence of the above gradient based update rule is given in Theorem \ref{thasli}.\\
Now the results for both leader and followers are used to solve the hierarchical game in the next subsection.

\subsection{The Main Hierarchical Game}

The errors, $ e_{i}$s, in \eqref{ei} and \eqref{en+1}, are used for updating critic NNs' weights with gradient descent approach:
\begin{equation}
E=\sum_{i=1}^{N+1}e_{i}^{T}e_{i}\quad \Rightarrow \quad \dot{\hat{\mathbf{W}}}_{v_{i}}=-\tau_{i}\frac{\partial E}{\partial \hat{\mathbf{W}}_{v_{i}}}
\end{equation}

The following facts can be concluded:
\begin{enumerate}\label{fact}
\item The NN approximation errors and their gradients are bounded:
$$ \Vert \epsilon_{v_i}\Vert <b_{\epsilon_{i}},\quad \Vert \nabla_{\mathbf{x}} \epsilon_{v_i}\Vert <b_{\epsilon_{x_i}} $$
\item The NN activation functions and their gradients are bounded:
$$ \Vert \boldsymbol{\phi}_{v_i}(\mathbf{x},\boldsymbol{\nu})\Vert <b_{\boldsymbol{\phi}_{i}},\quad \Vert \nabla_{\mathbf{x}} \boldsymbol{\phi}_{v_i}(\mathbf{x},\boldsymbol{\nu})\Vert <b_{\boldsymbol{\phi}_{i_x}}$$  $$
 \Vert \nabla_{\boldsymbol{\nu}}\nabla_{\mathbf{x}} \boldsymbol{\phi}_{v_i}(\mathbf{x},\boldsymbol{\nu})\Vert <b_{\boldsymbol{\phi}_{i_{x\boldsymbol{\nu}}}} $$
\item The ideal NN weights are bounded:
$$ \Vert \mathbf{W}_{v_i}\Vert < W_{i_{max}}, \quad \Vert W_{\nu}\Vert < W_{\nu_{max}} $$
\end{enumerate}
In the following theorem update rules and ultimate boundedness of errors are analyzed.
\begin{theorem} \label{thasli}
Consider the dynamical system \eqref{dynamic} with critic NNs \eqref{critic NN} and control inputs \eqref{u_hat} and \eqref{w_hat}. The tuning laws for the critic NNs are $($ for $ 1\leq i \leq N+1 ):$
\begin{align}
\dot{\hat{\mathbf{W}}}_{v_{i}}&=-\tau_{i}\boldsymbol{\mu}_{i}\Bigg( \boldsymbol{\eta}_{i}^{T}\hat{\mathbf{W}}_{v_{i}}+\frac{1}{4}\sum_{j=1}^{N}\hat{\mathbf{W}}_{a_{j}}^{T} C_{j}^{i}\hat{\mathbf{W}}_{a_{j}}+\hat{\boldsymbol{\nu}}S_{i}\hat{\boldsymbol{\nu}}\nonumber \\
&\quad -\frac{1}{4}\hat{\mathbf{W}}_{a_{i}}^{T}E_{i}\hat{\mathbf{W}}_{a_{i}}+Q_{i}(\mathbf{x})\Bigg)
\end{align}
where $ \boldsymbol{\eta}_{i}=\nabla \boldsymbol{\phi}_{v_i}(\mathbf{f}(\mathbf{x})+\sum_{j=1}^{N}g_{j}(\mathbf{x})\hat{\mathbf{u}}_{j}+p(\mathbf{x})\hat{\boldsymbol{\nu}}+h(\mathbf{x})\hat{\boldsymbol{\omega}} ) $, $ \rho_{i}=\boldsymbol{\eta}_{i}^{T}\boldsymbol{\eta}_{i}+1 $ , $ \bar{\boldsymbol{\eta}}_{i}=\frac{\boldsymbol{\eta}_{i}}{\rho_i} $ and $ \frac{\bar{\boldsymbol{\eta}}_{i}}{\rho_{i}}=\boldsymbol{\mu}_{i}$ .
The tuning laws for the leader NN and actor NNs are
\begin{align}
&\dot{\hat{W}}_{\nu}=-\varrho \Bigg\{\sum_{j=1}^{N}\bigg( \frac{\partial \hat{W}_{\nu}^{T}\boldsymbol{\phi}_{\nu}}{\partial \hat{W}_{\nu}}\bigg)\bigg( \frac{\partial^{2}\nabla \boldsymbol{\phi}_{v_{j}}}{\partial \hat{\boldsymbol{\nu}}^{2}}\bigg)^{T}\Bigg[-\frac{1}{4}\hat{A}_{j}^{N+1}\nonumber \\
&\qquad \times \hat{\Pi}^{T}S_{N+1}^{-2}+\frac{1}{8}\sum_{k=1}^{N}\hat{A}_{j}^{N+1}(\hat{A}_{k}^{N+1})^{T}\bigg(\frac{\partial \nabla \boldsymbol{\phi}_{v_k}}{\partial \hat{\boldsymbol{\nu}}}\bigg)S_{N+1}^{-2} \nonumber \\
&\qquad -\frac{1}{2}\hat{A}_{j}^{N+1}\boldsymbol{\phi}_{\nu}^{T}\hat{W}_{\nu}S_{N+1}^{-1}+\frac{1}{4}\hat{\mathbf{W}}_{a_j}\hat{\mathbf{W}}_{a_j}^{T}\nabla \boldsymbol{\phi}_{v_{j}}B_{j}^{N+1}\nonumber \\
&\qquad \times \hat{\Pi}^{T}S_{N+1}^{-2}+\frac{1}{8}\sum_{k=1}^{N}\hat{\mathbf{W}}_{a_j}\hat{\mathbf{W}}_{a_j}^{T}\nabla \boldsymbol{\phi}_{v_{j}}B_{j}^{N+1}\nonumber \\
&\qquad \times B_{k}^{N+1}\nabla \boldsymbol{\phi}_{v_{k}}^{T}\hat{\mathbf{W}}_{a_k}\hat{\mathbf{W}}_{a_k}^{T}\bigg( \frac{\partial \nabla \boldsymbol{\phi}_{v_{k}}}{\partial \hat{\boldsymbol{\nu}}}\bigg)S_{N+1}^{-2} \nonumber \\
&\qquad -\frac{1}{8}\sum_{k=1}^{N}\hat{A}_{j}^{N+1}B_{k}^{N+1}\nabla \boldsymbol{\phi}_{v_k}^{T}\hat{\mathbf{W}}_{a_k}\hat{\mathbf{W}}_{a_k}^{T}\bigg( \frac{\partial \nabla \boldsymbol{\phi}_{v_{k}}}{\partial \hat{\boldsymbol{\nu}}}\bigg)S_{N+1}^{-2} \nonumber \\
&\qquad +\frac{1}{2}\hat{\mathbf{W}}_{a_j}\hat{\mathbf{W}}_{a_j}^{T}\nabla \boldsymbol{\phi}_{v_{j}} B_{j}^{N+1}\boldsymbol{\phi}_{\nu}^{T}\hat{W}_{\nu}S_{N+1}^{-1}-\frac{1}{8}\sum_{k=1}^{N}\hat{\mathbf{W}}_{a_k}\nonumber \\
&\qquad \times \hat{\mathbf{W}}_{a_k}^{T}\nabla \boldsymbol{\phi}_{v_k}B_{k}^{N+1}(\hat{A}_{j}^{N+1})^{T}\bigg( \frac{\partial \nabla \boldsymbol{\phi}_{v_{j}}}{\partial \hat{\boldsymbol{\nu}}}\bigg)S_{N+1}^{-2}\Bigg] \nonumber \\
&\qquad +\sum_{j=1}^{N}\bigg( \frac{\partial \hat{W}_{\nu}^{T}\boldsymbol{\phi}_{\nu}}{\partial \hat{W}_{\nu}}\bigg)\bigg( \frac{\partial \nabla \boldsymbol{\phi}_{v_{j}}}{\partial \hat{\boldsymbol{\nu}}}\bigg)^{T}\Bigg[ \frac{1}{8}\sum_{k=1}^{N}\hat{\mathbf{W}}_{a_j}\hat{\mathbf{W}}_{a_j}^{T} \nonumber \\
&\qquad \times \bigg( \frac{\partial \nabla \boldsymbol{\phi}_{v_{j}}}{\partial \hat{\boldsymbol{\nu}}}\bigg)S_{N+1}^{-2}\bigg( \frac{\partial \nabla \boldsymbol{\phi}_{v_{k}}}{\partial \hat{\boldsymbol{\nu}}}\bigg)^{T}\hat{\mathbf{W}}_{a_k}\hat{\mathbf{W}}_{a_k}^{T}\nabla \boldsymbol{\phi}_{v_k}  \nonumber \\
&\qquad \times B_{j}^{N+1}B_{k}^{N+1}+\frac{1}{4}\hat{\mathbf{W}}_{a_j}\hat{\mathbf{W}}_{a_j}^{T}\bigg( \frac{\partial \nabla \boldsymbol{\phi}_{v_{j}}}{\partial \hat{\boldsymbol{\nu}}}\bigg)S_{N+1}^{-2}\hat{\Pi}\displaybreak[3] \nonumber \\
&\qquad +\frac{1}{2}\hat{\mathbf{W}}_{a_j}\hat{\mathbf{W}}_{a_j}^{T}\bigg( \frac{\partial \nabla \boldsymbol{\phi}_{v_{j}}}{\partial \hat{\boldsymbol{\nu}}}\bigg)S_{N+1}^{-1}\hat{W}_{\nu}^{T}\boldsymbol{\phi}_{\nu}B_{j}^{N+1} \nonumber \\
&\qquad -\frac{1}{8}\sum_{k=1}^{N}\hat{\mathbf{W}}_{a_k}\hat{\mathbf{W}}_{a_k}^{T}\bigg( \frac{\partial \nabla \boldsymbol{\phi}_{v_{k}}}{\partial \hat{\boldsymbol{\nu}}}\bigg)S_{N+1}^{-2}\bigg( \frac{\partial \nabla \boldsymbol{\phi}_{v_{j}}}{\partial \hat{\boldsymbol{\nu}}}\bigg)^{T}\nonumber \\
&\qquad \times \hat{A}_{j}^{N+1}B_{k}^{N+1}\Bigg]+2\boldsymbol{\phi}_{\nu}\boldsymbol{\phi}_{\nu}^{T}\hat{W}_{\nu} +\hat{\Pi}^{T}S_{N+1}^{-1}  \nonumber \\
&\qquad -\frac{1}{2}\sum_{j=1}^{N}\boldsymbol{\phi}_{\nu}(\hat{A}_{j}^{N+1})^{T}\bigg( \frac{\partial \nabla \boldsymbol{\phi}_{v_{j}}}{\partial \hat{\boldsymbol{\nu}}}\bigg)S_{N+1}^{-1}  \nonumber \\
&\qquad +\frac{1}{2}\sum_{j=1}^{N}\boldsymbol{\phi}_{\nu}B_{j}^{N+1}\nabla \boldsymbol{\phi}_{v_j}^{T}\hat{\mathbf{W}}_{a_j}\hat{\mathbf{W}}_{a_j}^{T}\bigg( \frac{\partial \nabla \boldsymbol{\phi}_{v_{j}}}{\partial \hat{\boldsymbol{\nu}}}\bigg)S_{N+1}^{-1}\Bigg\}
\end{align}
and
\begin{align}
 \dot{\hat{\mathbf{W}}}_{a_{i}}&=-\theta_{i}\Big[ \bigg(F_{2}^{i}\hat{\mathbf{W}}_{a_{i}}-F_{1}^{i}\bar{\boldsymbol{\eta}}_{i}^{T} \hat{\mathbf{W}}_{v_{i}}\bigg) -\frac{1}{4}\sum_{k=1}^{N+1}C_{i}^{k}\hat{\mathbf{W}}_{a_{i}}\nonumber \\
&\quad \times \boldsymbol{\mu}_{k}^{T}\hat{\mathbf{W}}_{v_{k}} +\frac{1}{4}E_{i}\hat{\mathbf{W}}_{a_{i}}\hat{\mathbf{W}}_{v_{i}} \Big], \quad 1\leq i\leq N
\end{align}
\begin{align}
\dot{\hat{\mathbf{W}}}_{a_{N+1}}&=-\theta_{N+1}\Big[ \bigg(F_{2}^{N+1}\hat{\mathbf{W}}_{a_{N+1}}-F_{1}^{N+1}\bar{\boldsymbol{\eta}}_{N+1}^{T} \hat{\mathbf{W}}_{v_{N+1}}\bigg)\nonumber \\
&\quad +\frac{1}{4}E_{N+1}\hat{\mathbf{W}}_{a_{N+1}}\boldsymbol{\mu}_{N+1}^{T}\hat{\mathbf{W}}_{v_{N+1}} \Big]
\end{align}
respectively.
 $ F_{1}^{i} $s and $ F_{2}^{i}$s are tuning parameters that can be selected as given in the proof of theorem in Appendix \ref{append}. In such condition there exists $ \kappa^{*} $ so that for $ \kappa>\kappa^{*} $ the closed loop system states, the critic and actor NN errors are UUB. Moreover, $ \hat{\mathbf{u}}_{i}$s, $\hat{\boldsymbol{\omega}}_{i}$s and $ \hat{\boldsymbol{\nu}}$ converge to the approximate Stackelberg-Nash-saddle equilibrium point of the game.
\end{theorem}
\begin{IEEEproof}
See Appendix \ref{append}.
\end{IEEEproof}
\begin{remark}
 Terms $ \big( \frac{\partial \hat{W}_{\nu}^{T}\boldsymbol{\phi}_{\nu}}{\partial \hat{W}_{\nu}}\big) $ and $ \boldsymbol{\phi}_{\nu} $ are not equal, because the first term is a gradient of a vector-valued function in $ \mathbb{R}^{\alpha} $ with respect to a matrix in $ \mathbb{R}^{\mu \times \alpha} $ and hence is in $ \mathbb{R}^{\mu \times \alpha \times \alpha} $ while the second term is a vector in $ \mathbb{R}^{\mu} $. In some especial cases these two terms become equal; e.g., $ \alpha =1 $ .
\end{remark}

\section{Simulation}

In this section the proposed algorithm is implemented on a hierarchical one-leader-two-followers game with nonlinear dynamic. Disturbance is considered as a uniform noise bounded in $ [-0.5 , 0.5]$ . following is the system dynamics:
\begin{equation}
\dot{\mathbf{x}}=\mathbf{f}(\mathbf{x})+\mathbf{g}_{1}(\mathbf{x})u_1+\mathbf{g}_{2}(\mathbf{x})u_2+\mathbf{p}(\mathbf{x})\nu +\mathbf{h}(\mathbf{x})\omega
\end{equation}
where $ \mathbf{f}(\mathbf{x})=[f_{1}(\mathbf{x}) \quad f_{2}(\mathbf{x})]^T $ and
\begin{align*}
f_{1}(\mathbf{x})&= x_2 \nonumber \\
f_{2}(\mathbf{x})&=-x_2+0.5x_1+0.25x_2(\cos (2x_1)+2)^2 \nonumber \\
&\qquad +0.25x_2(\sin (4x_1^2)+2)^2
\end{align*}
\begin{equation*}
\mathbf{g}_{1}(\mathbf{x})=\begin{bmatrix}
0\\
\cos (2x_1)+2
\end{bmatrix} ,\qquad \mathbf{g}_{1}(\mathbf{x})=\begin{bmatrix}
0\\
\sin (4x_1^2)+2
\end{bmatrix}
\end{equation*}
and
\begin{equation*}
\mathbf{p}(\mathbf{x})=\begin{bmatrix}
1\\
\cos (x_1^2)+4
\end{bmatrix} , \qquad \mathbf{h}(\mathbf{x})=\begin{bmatrix}
\sin (5x_1)+0.1\\
\cos (2x_1^2)+0.2
\end{bmatrix}
\end{equation*}
The parameters of performance index functions are $ Q_{1}(x)=2x_{1}^2+x_{2}^{2} $, $ Q_{2}(x)=x_{1}^2 +4x_{2}^{2} $, $ Q_{3}(x)=x_{1}^{4}+2x_{2}^{2} $, $ S_{1}=4 $, $ S_{2}=2 $, $ S_{3}=20 $ (since in real situations the effect of leader is more than the effect of followers, $S_3$ is more than $S_2$ and $S_1$), $ R_{11}=4 $, $ R_{12}=1 $, $ R_{21}=1 $, $ R_{22}=2 $, $ R_{31}=1 $, $ R_{32}=1 $ and  the disturbance attenuation $ \gamma^2 =0.6 $. The initial state $ x_{1}(0)=x_{2}(0)=0.01 $. Initial values of NN weights, $ \hat{\mathbf{W}}_{v_i} $s, $ \hat{\mathbf{W}}_{a_i} $s and $ \hat{\mathbf{W}}_{\nu} $ for $ 1\leq i\leq 3 $, are randomly selected in the interval $ (0, 1) $. Activation functions are $ \boldsymbol{\phi}_{v_1}=\boldsymbol{\phi}_{v_2}=[x_1^2, x_2^2, x_{1}x_{2}, x_{1}\nu, x_{2}\nu ]^{T} $, $ \boldsymbol{\phi}_{v_3}=[x_{1}^2, x_{2}^{2}, x_{1}^{4}, x_{2}^{4}, x_{1}^{2}x_{2}^{2}]^{T} $ and $ \boldsymbol{\phi}_{\nu}=[x_{1}, x_{2}, x_{1}^{2}, x_{2}^{2}, x_{1}x_{2}]^{T} $. Learning rates are $ \tau_{1}=\tau_{2}=0.7 $ , $ \tau_{3}=0.6 $ , $ \theta_{1}=\theta_{2}=\theta_{3}=0.7 $ and $ \varrho =0.6 $. Tuning parameters $ F_{2}^{1}=F_{2}^{2}=F_{2}^{3}=200 $ and $ F_{1}^{1}=F_{1}^{2}=F_{1}^{3}=100\times [1, 1, 1, 1, 1]^{T} $ .
\begin{figure}
\centering
\begin{subfigure}[b]{0.45\textwidth}
\includegraphics[scale=0.45]{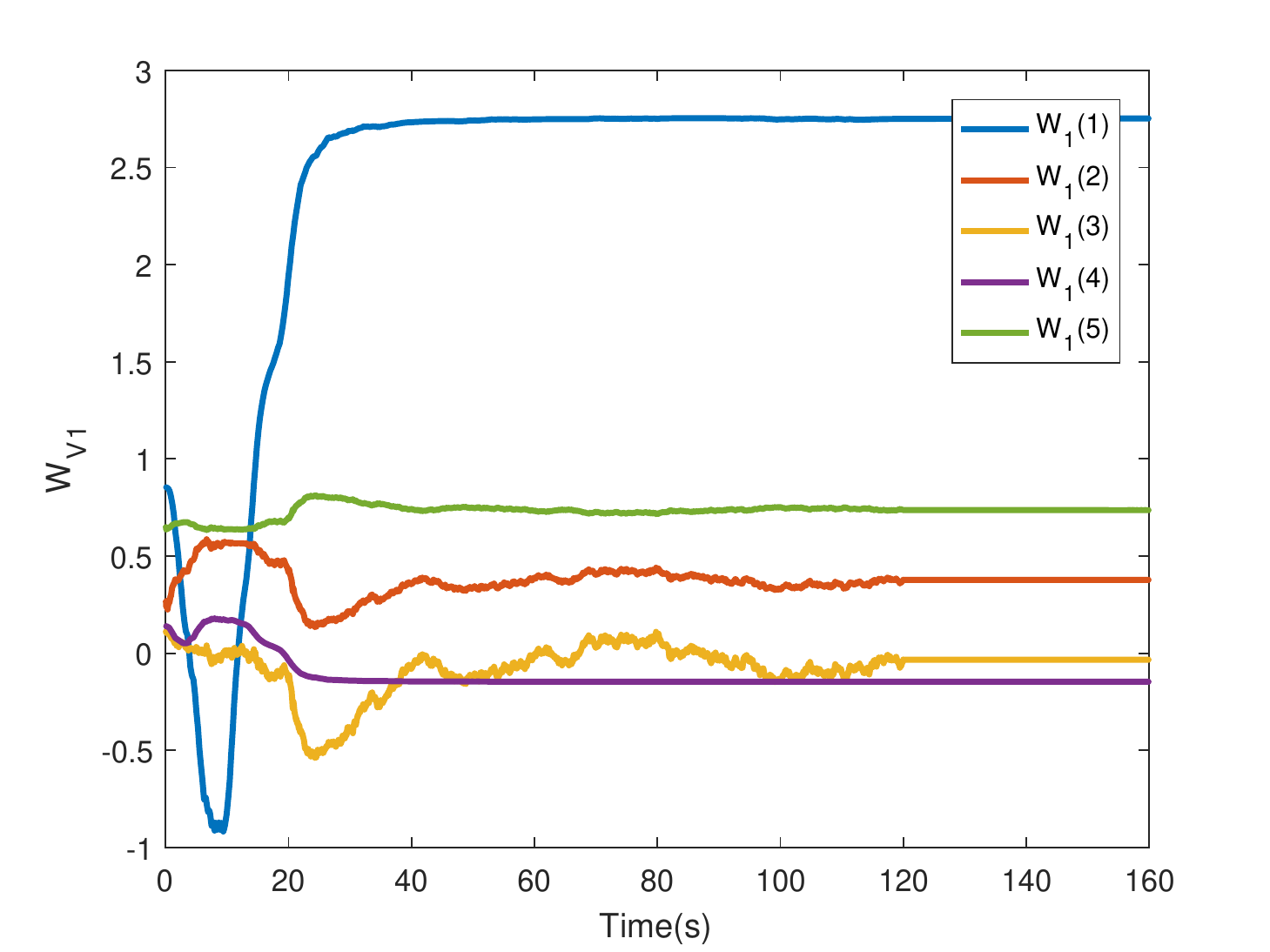}
\caption{First follower.}
\label{figv1}
\end{subfigure}
\begin{subfigure}[b]{0.45\textwidth}
\includegraphics[scale=0.45]{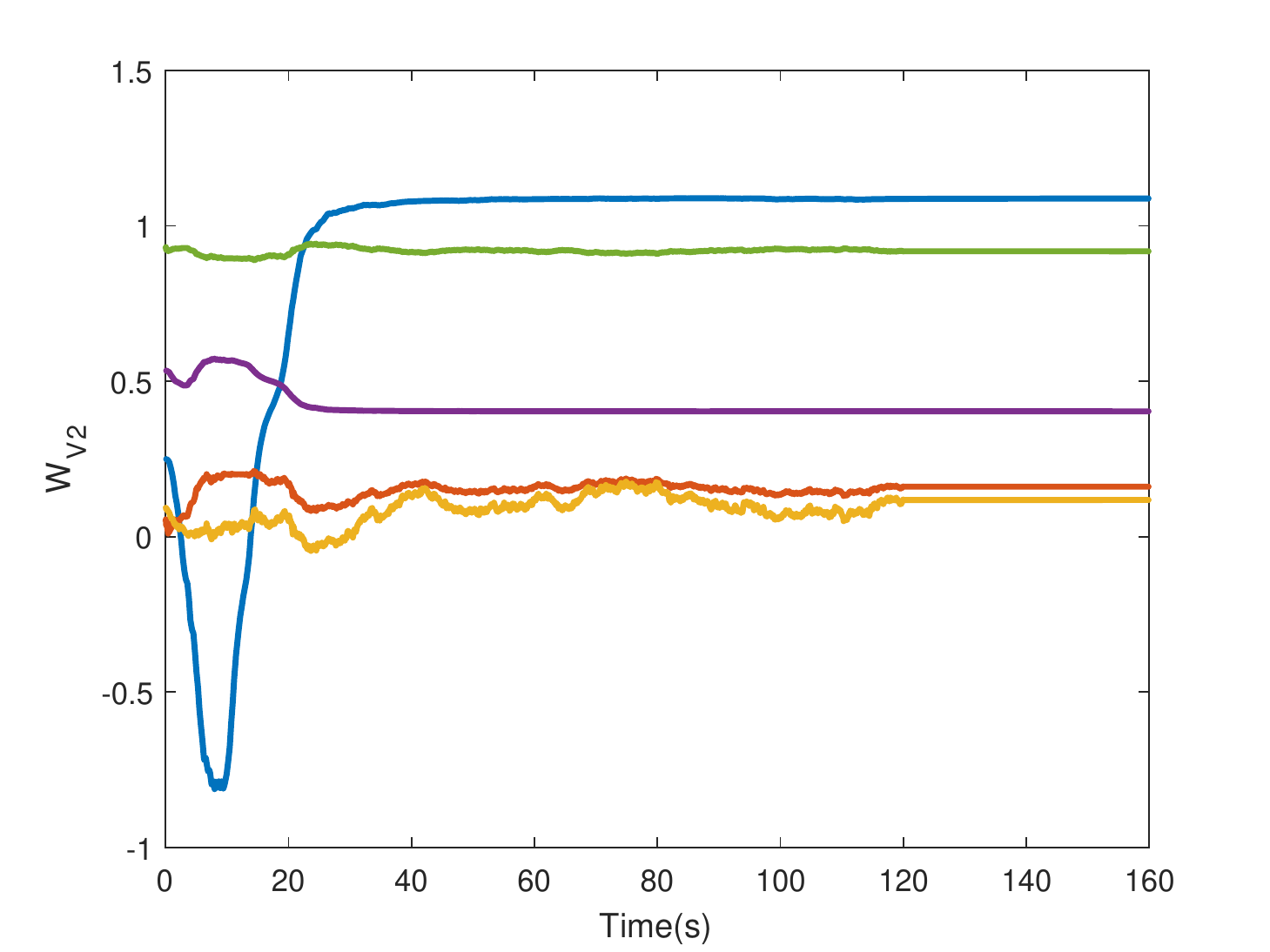}
\caption{Second follower.}
\label{figv2}
\end{subfigure}
\begin{subfigure}[b]{0.45\textwidth}
\includegraphics[scale=0.45]{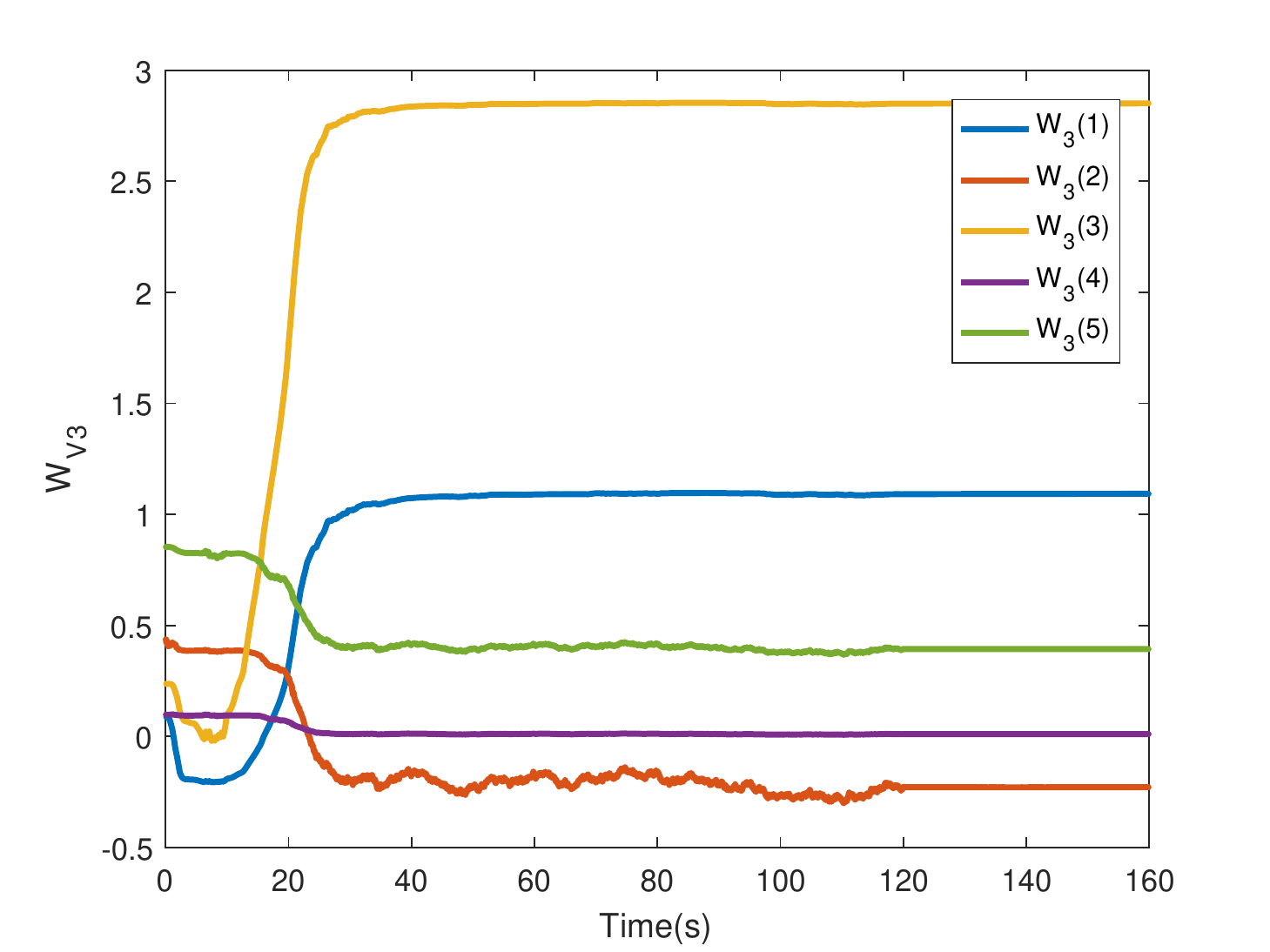}
\caption{Leader.}
\label{figv3}
\end{subfigure}
\caption{Convergence of the critic NNs.}
\label{CCNN}
\end{figure}
Also a white Gaussian probing noise is added for satisfying persistence of exitation condition. After convergence at about 80 seconds one could get $ \mathbf{W}_{v_1}=\mathbf{W}_{a_1}=[2.752, 0.3777, -0.0334, -0.1469, 0.7363]^{T} $, $ \mathbf{W}_{v_2}=\mathbf{W}_{a_2}=[1.0876, 0.1607, 0.1180, 0.4029, 0.9177]^{T} $, $ \mathbf{W}_{v_3}=\mathbf{W}_{a_3}=[1.0929, -0.2286, 2.8509, 0.0107, 0.3939]^{T} $ and $ \mathbf{W}_{\nu}=[-0.4446, -0.0240, 0.0643, 0.5727, 0.5827, 0.2225]^{T} $ .System states are presented in Fig. \ref{figs}.
%
\begin{figure}[!ht]
\begin{center}
\includegraphics[scale=0.45]{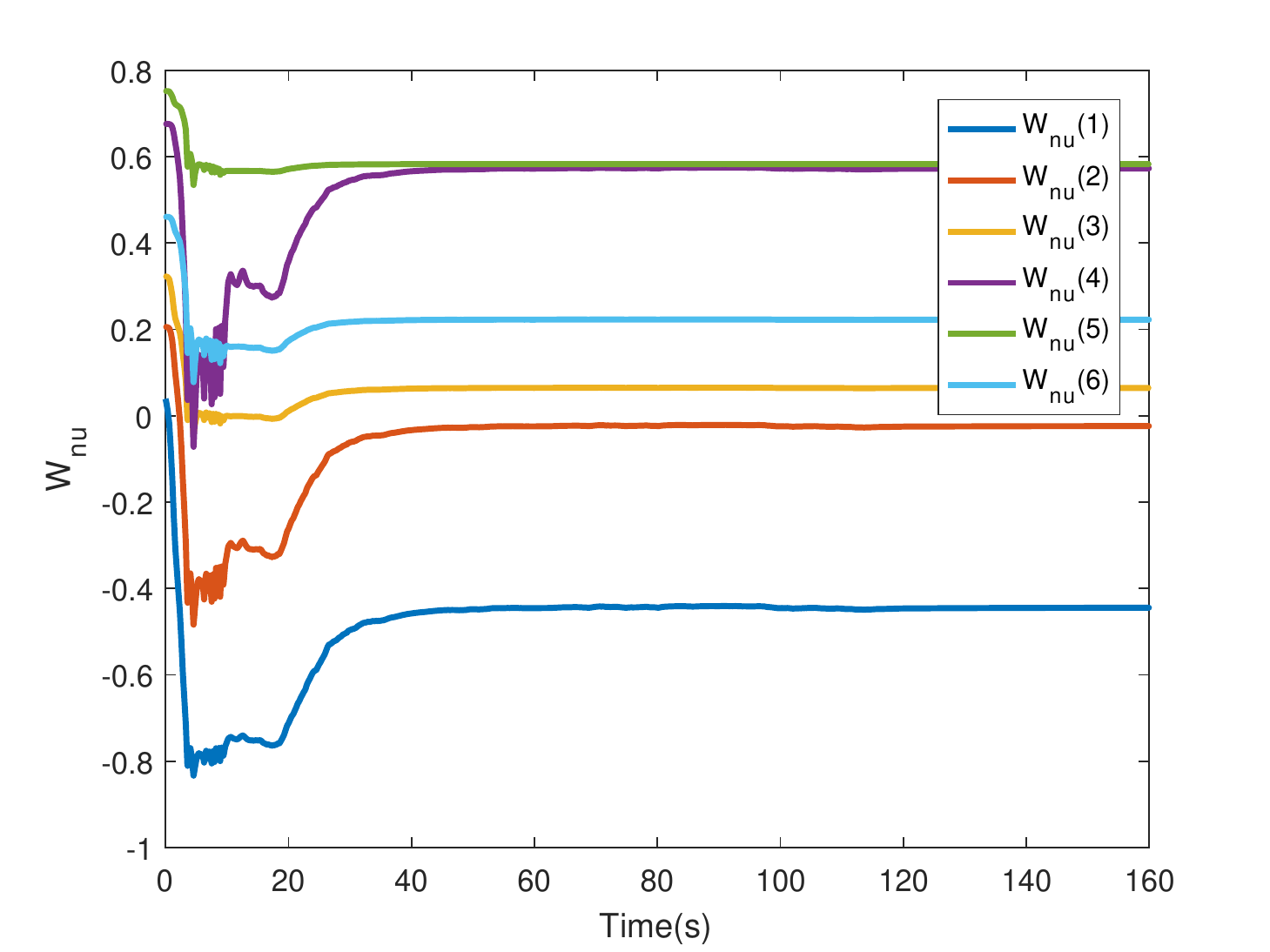}
\end{center}
\caption{Convergence of the leader NN.}
\label{fignu}
\end{figure}

\begin{figure}[!ht]
\begin{center}
\includegraphics[scale=0.45]{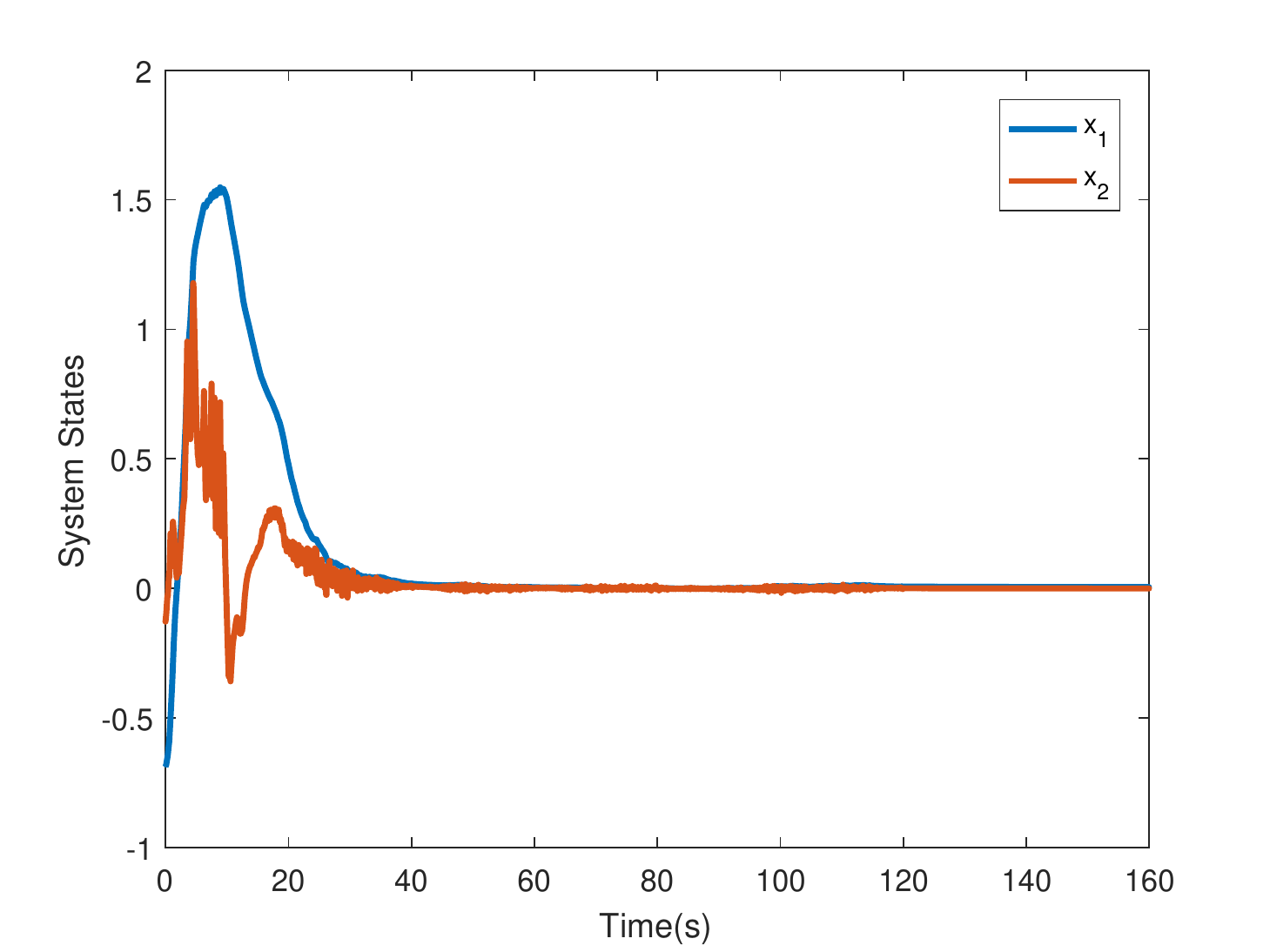}
\end{center}
\caption{Evolution of the system states.}
\label{figs}
\end{figure}

\section{Conclusion}

In this paper an on-line ADP-based method is developed for solving a class of hierarchical one-leader-multi-followers nonlinear differential games. The game discussed here was made up of both zero-sum and nonzero-sum games. In the proposed algorithm the value functions approximated with NNs and the approximations improved by gradient descent. Also the actor NN weights were updated using adaptive rules. Using Lyapunov theory, it was showed that the NN weights and system states are UUB. It was assumed that the game had one equilibrium and the condition which satisfied this assumption was derived using Nemytskii operator.

As a future work of this study, the effect of model parameter uncertainties  on the convergence of the on-line ADP can be investigated.


%

\appendix[Proof of theorem 7]\label{append}

\begin{IEEEproof}
The candidate Lyapunov function is defined as follows
\begin{align}
L(t) &=\sum_{i=1}^{N+1}V_{i}(\mathbf{x})+\sum_{i=1}^{N+1}\frac{1}{2}\tilde{\mathbf{W}}_{v_{i}}^{T}\tau_{i}^{-1}\tilde{\mathbf{W}}_{v_{i}}\nonumber \\
&\qquad +\sum_{i=1}^{N+1}\frac{1}{2}\tilde{\mathbf{W}}_{a_{i}}^{T}\theta_{i}^{-1}\tilde{\mathbf{W}}_{a_{i}} \label{lyapunov}
\end{align}
where $ \mathbf{W}_{v_{i}}-\hat{\mathbf{W}}_{v_{i}}=\tilde{\mathbf{W}}_{v_{i}} $ and
$ \mathbf{W}_{v_{i}}-\hat{\mathbf{W}}_{a_{i}}=\tilde{\mathbf{W}}_{a_{i}}$.
Time derivative of \eqref{lyapunov} is:
\begin{equation}
\dot{L}(t) =\sum_{i=1}^{N+1}\dot{V}_{i}(\mathbf{x}) +\sum_{i=1}^{N+1}\tilde{\mathbf{W}}_{v_{i}}^{T}\tau_{i}^{-1}\dot{\tilde{\mathbf{W}}}_{v_{i}} +\sum_{i=1}^{N+1}\tilde{\mathbf{W}}_{a_{i}}^{T}\theta_{i}^{-1}\dot{\tilde{\mathbf{W}}}_{a_{i}} \label{ldot}
\end{equation}
Since the time derivative of value functions $ \dot{V}_{i} = (\nabla V_{i})^{T}\dot{\mathbf{x}}=(\mathbf{W}_{v_{i}}^{T}\nabla \boldsymbol{\phi}_{v_{i}}+\nabla \epsilon_{i}^{T})\dot{\mathbf{x}} $ , one can get:
\begin{align}
\dot{V}_{i}&=\mathbf{W}_{v_{i}}^{T}\Bigg( \nabla \boldsymbol{\phi}_{v_{i}}\mathbf{f}(\mathbf{x})-\frac{1}{2}\sum_{j=1}^{N}D_{i}^{j}\hat{\mathbf{W}}_{a_{j}}+\frac{1}{2}E_{i}\hat{\mathbf{W}}_{a_{i}}\nonumber \\
&\quad +\nabla \boldsymbol{\phi}_{v_{i}}p(\mathbf{x})\hat{\boldsymbol{\nu}} \Bigg) \pm \frac{1}{2}\sum_{j=1}^{N}\mathbf{W}_{v_{i}}^{T}D_{i}^{j}\mathbf{W}_{a_{j}} \nonumber \\
&\qquad \pm \frac{1}{2}\mathbf{W}_{v_{i}}^{T}E_{i}\mathbf{W}_{a_{i}}+\dot{\epsilon}_{i} \displaybreak[3] \nonumber \\
&=\mathbf{W}_{v_{i}}^{T}\boldsymbol{\sigma}_{i}+\frac{1}{2}\mathbf{W}_{v_{i}}^{T}\sum_{j=1}^{N}D_{i}^{j}\tilde{\mathbf{W}}_{a_{j}}-\frac{1}{2}\mathbf{W}_{v_{i}}^{T}E_{i}\tilde{\mathbf{W}}_{a_{i}} \nonumber \\
&\qquad +\mathbf{W}_{v_{i}}^{T}\nabla \boldsymbol{\phi}_{v_{i}}p(\mathbf{x})(\hat{\boldsymbol{\nu}}-\check{\boldsymbol{\nu}})+\dot{\epsilon}_{i} \label{vidot}
\end{align}
where
\begin{align}
\dot{\epsilon}_{i}&=\nabla \epsilon_{i}^{T}\Bigg( \mathbf{f}(\mathbf{x})-\frac{1}{2}\sum_{j=1}^{N}g_{j}(\mathbf{x})R_{jj}^{-1}g_{j}^{T}(\mathbf{x})\nabla \boldsymbol{\phi}_{v_{j}}^{T}\hat{\mathbf{W}}_{a_{j}}\nonumber \\
&\qquad +\frac{\gamma^{-2}}{2}h(\mathbf{x})h^{T}(\mathbf{x})\nabla \boldsymbol{\phi}_{v_{i}}^{T}\hat{\mathbf{W}}_{a_{i}}+p(\mathbf{x})\hat{\boldsymbol{\nu}}\Bigg) \label{epsdot} \\
\boldsymbol{\sigma}_{i} &=\nabla \boldsymbol{\phi}_{v_{i}}\Bigg( \mathbf{f}(\mathbf{x})-\frac{1}{2}\sum_{j=1}^{N} g_{j}(\mathbf{x})R_{jj}^{-1}g_{j}^{T}(\mathbf{x})\nabla \boldsymbol{\phi}_{v_{j}}^{T}\mathbf{W}_{v_{j}}\nonumber \\
&\qquad +p(\mathbf{x})\check{\boldsymbol{\nu}}+\frac{\gamma^{-2}}{2}h(\mathbf{x})h^{T}(\mathbf{x})\nabla \boldsymbol{\phi}_{v_{i}}^{T}\mathbf{W}_{v_{i}} \Bigg) \label{sigmai}
\end{align}
From \eqref{ephji_main} it is concluded that:
\begin{align}
\mathbf{W}_{v_{i}}^{T}\boldsymbol{\sigma}_{i}&=-Q_{i}(\mathbf{x})-\frac{1}{4}\sum_{j=1}^{N}\mathbf{W}_{v_{j}}^{T}C_{j}^{i}\mathbf{W}_{v_{j}}\displaybreak[3]\nonumber \\
&\qquad +\frac{1}{4}\mathbf{W}_{v_{i}}^{T}E_{i}\mathbf{W}_{v_{i}} -\check{\boldsymbol{\nu}}^{T}S_{i}\check{\boldsymbol{\nu}}+\epsilon_{HJ_{i}} \label{wsig}
\end{align}
Now the second part of \eqref{ldot} is calculated according to \eqref{ei} and gradient descent method.
\begin{align}
\tilde{\mathbf{W}}_{v_{i}}^{T}\tau_{i}^{-1}\dot{\tilde{\mathbf{W}}}_{v_{i}} &= \tilde{\mathbf{W}}_{v_{i}}^{T}\frac{\boldsymbol{\eta}_{i}}{(\boldsymbol{\eta}_{i}^{T}\boldsymbol{\eta}_{i}+1)^{2}}\Bigg( \boldsymbol{\eta}_{i}^{T}\hat{\mathbf{W}}_{v_{i}}+Q_{i}(\mathbf{x})\nonumber \\
&\qquad +\sum_{j=1}^{N}\hat{\mathbf{u}}_{j}^{T}R_{ij}\hat{\mathbf{u}}_{j}+ \hat{\boldsymbol{\nu}}S_{i}\hat{\boldsymbol{\nu}}-\gamma^{2}\hat{\boldsymbol{\omega}}^{T}\hat{\boldsymbol{\omega}}\Bigg)\label{47}
\end{align}
Now the terms in \eqref{47} which are in parentheses can be rewritten as
\begin{align}
& Q_{i}(\mathbf{x})+ \frac{1}{4}\sum_{j=1}^{N}\big( \mathbf{W}_{v_{j}}-\tilde{\mathbf{W}}_{a_{j}}\big)^{T}C_{j}^{i}(\mathbf{W}_{v_{j}}-\tilde{\mathbf{W}}_{a_{j}}) \nonumber \\
& -\frac{1}{4}(\mathbf{W}_{v_{i}}-\tilde{\mathbf{W}}_{a_{i}})^{T}E_{i}(\mathbf{W}_{v_{i}}-\tilde{\mathbf{W}}_{a_{i}})\nonumber \\
&+(\mathbf{W}_{v_{i}}-\tilde{\mathbf{W}}_{v_{i}})^{T}\boldsymbol{\eta}_{i}+\hat{\boldsymbol{\nu}}^{T}S_{i}\hat{\boldsymbol{\nu}} \label{above1}
\end{align}
According to \eqref{sigmai} and \eqref{wsig}, \eqref{above1} becomes
\begin{align}
& -\tilde{\mathbf{W}}_{v_{i}}^{T}\boldsymbol{\eta}_{i}+\epsilon_{HJ_{i}}+\frac{1}{4}\sum_{j=1}^{N}\tilde{\mathbf{W}}_{a_{j}}C_{j}^{i}\tilde{\mathbf{W}}_{a_{j}}+\boldsymbol{\phi}_{\nu}^{T}\hat{W}_{\nu}S_{i}  \nonumber \\
&\qquad \times \hat{W}_{\nu}^{T}\boldsymbol{\phi}_{\nu} -\frac{1}{4}\tilde{\mathbf{W}}_{a_{i}}^{T}E_{i}\tilde{\mathbf{W}}_{a_{i}}-\boldsymbol{\phi}_{\nu}^{T}W_{\nu}S_{i}W_{\nu}^{T}\boldsymbol{\phi}_{\nu} \displaybreak[3] \nonumber \\
&\qquad  -\frac{1}{2}\sum_{\substack{j=1 \\ j\neq i}}^{N}\mathbf{W}_{v_{j}}^{T}C_{j}^{i}\tilde{\mathbf{W}}_{a_{j}} \nonumber \\
&\qquad +\mathbf{W}_{v_{i}}^{T}\bigg(- \nabla \boldsymbol{\phi}_{v_{i}}p(\mathbf{x})\tilde{W}_{\nu}^{T}\boldsymbol{\phi}_{\nu}+\frac{1}{2}\sum_{\substack{j=1 \\ j\neq i}}^{N}D_{j}^{i}\tilde{\mathbf{W}}_{a_{j}} \bigg)
\end{align}
Thus
\begin{align}
& \tilde{\mathbf{W}}_{v_{i}}^{T}\tau_{i}^{-1}\dot{\tilde{\mathbf{W}}}_{v_{i}}=\tilde{\mathbf{W}}_{v_{i}}^{T}\frac{\boldsymbol{\eta}_{i}}{(\boldsymbol{\eta}_{i}^{T}\boldsymbol{\eta}_{i}+1)^{2}}\Bigg( -\tilde{\mathbf{W}}_{v_{i}}^{T}\boldsymbol{\eta}_{i}\nonumber \\
&\qquad +\frac{1}{4}\sum_{j=1}^{N}\tilde{\mathbf{W}}_{a_{j}}^{T}C_{j}^{i}\tilde{\mathbf{W}}_{a_{j}}-\frac{1}{2}\sum_{\substack{j=1 \\ j\neq i}}^{N}\mathbf{W}_{v_{j}}^{T}C_{j}^{i}\tilde{\mathbf{W}}_{a_{j}} \nonumber \\
&\qquad +\mathbf{W}_{v_{i}}^{T}\bigg(- \nabla \boldsymbol{\phi}_{v_{i}}p(\mathbf{x})\tilde{W}_{\nu}^{T}\boldsymbol{\phi}_{\nu}g_{j}(\mathbf{x})R_{jj}^{-1}\nonumber \\
&\qquad +\frac{1}{2}\sum_{\substack{j=1 \\ j\neq i}}^{N}D_{i}^{j}\tilde{\mathbf{W}}_{a_{j}}\bigg) -2\boldsymbol{\phi}_{\nu}^{T}\tilde{W}_{\nu}S_{i}W_{\nu}^{T}\boldsymbol{\phi}_{\nu}\displaybreak[3] \nonumber \\
&\qquad +\epsilon_{HJ_{i}}-\frac{\gamma^{-2}}{4}\tilde{\mathbf{W}}_{a_{i}}^{T}E_{i}\tilde{\mathbf{W}}_{a_{i}}+ \boldsymbol{\phi}_{\nu}^{T}\tilde{W}_{\nu}S_{i}\tilde{W}_{\nu}^{T}\boldsymbol{\phi}_{\nu}\Bigg) \label{wtildvi}
\end{align}
Now according to \eqref{vidot}, \eqref{epsdot}, \eqref{wsig}, and \eqref{wtildvi}, \eqref{ldot} is rewritten as:
\begin{align}
& \dot{L}(\mathbf{x}) = \sum_{i=1}^{N+1}\Bigg( -Q_{i}(\mathbf{x})+\epsilon_{HJ_{i}}-\frac{1}{4}\sum_{j=1}^{N}\mathbf{W}_{v_{j}}^{T}C_{j}^{i}\mathbf{W}_{v_{j}}\nonumber \\
&\qquad +\frac{1}{4}\sum_{j=1}^{N} \tilde{\mathbf{W}}_{a_{j}}^{T}C_{j}^{i}\tilde{\mathbf{W}}_{a_{j}}\boldsymbol{\mu}_{i}^{T}\tilde{\mathbf{W}}_{v_{i}} \nonumber \\
&\qquad +\frac{1}{4}\mathbf{W}_{v_{i}}^{T}E_{i}\mathbf{W}_{v_{i}}- \tilde{\mathbf{W}}_{v_{i}}^{T}\bar{\boldsymbol{\eta}}_{i}\bar{\boldsymbol{\eta}}_{i}^{T}\tilde{\mathbf{W}}_{v_{i}} -\frac{1}{2}\sum_{\substack{j=1 \\ j\neq i}}^{N}\tilde{\mathbf{W}}_{a_{j}}^{T} \displaybreak[3] \nonumber \\
&\qquad \times C_{j}^{i} \mathbf{W}_{v_{j}}\boldsymbol{\mu}_{i}^{T}\tilde{\mathbf{W}}_{v_{i}}+\tilde{\mathbf{W}}_{a_{i}}^{T}\theta_{i}^{-1}\dot{\tilde{\mathbf{W}}}_{a_{i}} \nonumber \\
&\qquad +\frac{1}{2}\sum_{\substack{j=1 \\ j\neq i}}^{N}\tilde{\mathbf{W}}_{a_{j}}^{T}D_{i}^{j}\mathbf{W}_{v_{i}}\boldsymbol{\mu}_{i}^{T}\tilde{\mathbf{W}}_{v_{i}}+\frac{1}{2}\mathbf{W}_{v_{i}}^{T}\sum_{j=1}^{N}D_{i}^{j} \nonumber \\
&\qquad \times \tilde{\mathbf{W}}_{a_{j}} -\frac{1}{2}\mathbf{W}_{v_{i}}^{T}E_{i}\tilde{\mathbf{W}}_{a_{i}}+\tilde{\mathbf{W}}_{v_{i}}^{T}\boldsymbol{\mu}_{i}\epsilon_{HJ_{i}} \nonumber \\
&\qquad -\mathbf{W}_{v_{i}}^{T}\nabla \boldsymbol{\phi}_{v_{i}} p(\mathbf{x})\tilde{W}_{\nu}^{T}\boldsymbol{\phi}_{\nu}+ \boldsymbol{\phi}_{\nu}^{T}\tilde{W}_{\nu}S_{i}\tilde{W}_{\nu}^{T}\boldsymbol{\phi}_{\nu} \nonumber \\
&\qquad \times \boldsymbol{\mu}_{i}^{T}\tilde{\mathbf{W}}_{v_{i}} -\tilde{\mathbf{W}}_{v_{i}}^{T}\boldsymbol{\mu}_{i}\mathbf{W}_{v_{i}}^{T}\nabla \boldsymbol{\phi}_{v_{i}} p(\mathbf{x})\tilde{W}_{\nu}^{T}\boldsymbol{\phi}_{\nu}
\nonumber \\
&\qquad -\boldsymbol{\phi}_{\nu}^{T}W_{\nu}S_{i}W_{\nu}^{T}\boldsymbol{\phi}_{\nu}- 2\boldsymbol{\phi}_{\nu}^{T}\tilde{W}_{\nu}S_{i}W_{\nu}^{T}\boldsymbol{\phi}_{\nu} \nonumber \\
&\qquad \times \boldsymbol{\mu}_{i}^{T}\tilde{\mathbf{W}}_{v_{i}} -\frac{1}{4}\tilde{\mathbf{W}}_{a_{i}}^{T}E_{i}\tilde{\mathbf{W}}_{a_{i}}\boldsymbol{\mu}_{i}^{T}\tilde{\mathbf{W}}_{v_{i}}  \nonumber \\
&\qquad +\nabla \epsilon_{i}^{T}\Big( \mathbf{f}(\mathbf{x})-\frac{1}{2}\sum_{j=1}^{N}g_{j}(\mathbf{x})R_{jj}^{-1}g_{j}^{T}(\mathbf{x})\nabla \boldsymbol{\phi}_{v_{j}}^{T}\hat{\mathbf{W}}_{a_{j}}\nonumber \\
&\qquad +\frac{\gamma^{-2}}{2}h(\mathbf{x})h^{T}(\mathbf{x})\nabla \boldsymbol{\phi}_{v_{i}}^{T}\hat{\mathbf{W}}_{a_{i}}+p(\mathbf{x})\hat{\boldsymbol{\nu}}\Big) \Bigg)\label{123}
\end{align}
Expanding some terms in \eqref{123} yields:
\begin{align}
\dot{L}(\mathbf{x}) &= \sum_{i=1}^{N+1}\Bigg( -Q_{i}(\mathbf{x})+\epsilon_{HJ_{i}}-\frac{1}{4}\sum_{j=1}^{N}\mathbf{W}_{v_{j}}^{T}C_{j}^{i}\mathbf{W}_{v_{j}} \nonumber \\
&\qquad -\frac{1}{4}\sum_{j=1}^{N} \tilde{\mathbf{W}}_{a_{j}}^{T}C_{j}^{i}\mathbf{W}_{v_{j}}\boldsymbol{\mu}_{i}^{T}\mathbf{W}_{v_{i}}+\nabla \epsilon_{i}^{T}\mathbf{f}(\mathbf{x}) \nonumber \\
&\qquad +\frac{1}{4}\sum_{j=1}^{N} \tilde{\mathbf{W}}_{a_{j}}^{T}C_{j}^{i}\mathbf{W}_{v_{j}}\boldsymbol{\mu}_{i}^{T}\tilde{\mathbf{W}}_{v_{i}}-\tilde{\mathbf{W}}_{v_{i}}^{T}\bar{\boldsymbol{\eta}}_{i}\bar{\boldsymbol{\eta}}_{i}^{T}\tilde{\mathbf{W}}_{v_{i}}  \nonumber \\
&\qquad -\frac{1}{2}\sum_{\substack{j=1 \\ j\neq i}}^{N}\tilde{\mathbf{W}}_{a_{j}}^{T}C_{j}^{i}\mathbf{W}_{v_{j}}\boldsymbol{\mu}_{i}^{T}\tilde{\mathbf{W}}_{v_{i}}  \nonumber \\
&\qquad +\frac{1}{2}\sum_{\substack{j=1 \\ j\neq i}}^{N}\tilde{\mathbf{W}}_{a_{j}}^{T}D_{i}^{j}\mathbf{W}_{v_{i}}\boldsymbol{\mu}_{i}^{T}\tilde{\mathbf{W}}_{v_{i}}+\frac{1}{2}\mathbf{W}_{v_{i}}^{T}\sum_{j=1}^{N}D_{i}^{j}\nonumber \\
&\qquad \times \tilde{\mathbf{W}}_{a_{j}} -\frac{1}{2}\mathbf{W}_{v_{i}}^{T}E_{i}\tilde{\mathbf{W}}_{a_{i}}+\tilde{\mathbf{W}}_{v_{i}}^{T}\boldsymbol{\mu}_{i}\epsilon_{HJ_{i}} \nonumber \\
&\qquad -\mathbf{W}_{v_{i}}^{T}\nabla \boldsymbol{\phi}_{v_{i}} p(\mathbf{x})\tilde{W}_{\nu}^{T}\boldsymbol{\phi}_{\nu}+ \boldsymbol{\phi}_{\nu}^{T}\tilde{W}_{\nu}S_{i}\tilde{W}_{\nu}^{T}\boldsymbol{\phi}_{\nu}\nonumber \\
&\qquad \times \boldsymbol{\mu}_{i}^{T}\tilde{\mathbf{W}}_{v_{i}}-\tilde{\mathbf{W}}_{v_{i}}^{T}\boldsymbol{\mu}_{i}\mathbf{W}_{v_{i}}^{T}\nabla \boldsymbol{\phi}_{v_{i}} p(\mathbf{x})\tilde{W}_{\nu}^{T}\boldsymbol{\phi}_{\nu}\nonumber \\
&\qquad +\frac{1}{4}\tilde{\mathbf{W}}_{a_{i}}^{T}E_{i}\mathbf{W}_{v_{i}}\boldsymbol{\mu}_{i}^{T}\mathbf{W}_{v_{i}} -\boldsymbol{\phi}_{\nu}^{T}W_{\nu}S_{i}W_{\nu}^{T}\boldsymbol{\phi}_{\nu}\nonumber \\
&\qquad +\frac{1}{4}\sum_{j=1}^{N} \tilde{\mathbf{W}}_{a_{j}}^{T}C_{j}^{i}\tilde{\mathbf{W}}_{a_{j}}\boldsymbol{\mu}_{i}^{T}\mathbf{W}_{v_{i}}-\frac{1}{4}\tilde{\mathbf{W}}_{a_{i}}^{T}E_{i}\tilde{\mathbf{W}}_{a_{i}}\nonumber \\
&\qquad \times \boldsymbol{\mu}_{i}^{T}\mathbf{W}_{v_{i}} -2\boldsymbol{\phi}_{\nu}^{T}\tilde{W}_{\nu}S_{i}W_{\nu}^{T}\boldsymbol{\phi}_{\nu}\boldsymbol{\mu}_{i}^{T}\tilde{\mathbf{W}}_{v_{i}} \nonumber \\
&\qquad +\frac{1}{4}\mathbf{W}_{v_{i}}^{T}E_{i}\mathbf{W}_{v_{i}} -\frac{1}{4}\tilde{\mathbf{W}}_{a_{i}}^{T}E_{i}\mathbf{W}_{v_{i}}\boldsymbol{\mu}_{i}^{T}\tilde{\mathbf{W}}_{v_{i}} \nonumber \\
&\qquad +\nabla \epsilon_{i}^{T}\Big( -\frac{1}{2}\sum_{j=1}^{N}g_{j}(\mathbf{x})R_{jj}^{-1}g_{j}^{T}(\mathbf{x})\nabla \boldsymbol{\phi}_{v_{j}}^{T}\mathbf{W}_{v_{j}}\nonumber \\
&\qquad +\frac{\gamma^{-2}}{2}h(\mathbf{x})h^{T}(\mathbf{x})\nabla \boldsymbol{\phi}_{v_{i}}^{T}\mathbf{W}_{v_{i}}+p(\mathbf{x})\check{\boldsymbol{\nu}}\Big) \nonumber \\
&\qquad -\nabla \epsilon_{i}^{T}\Big( -\frac{1}{2}\sum_{j=1}^{N}g_{j}(\mathbf{x})R_{jj}^{-1}g_{j}^{T}(\mathbf{x})\nabla \boldsymbol{\phi}_{v_{j}}^{T}\tilde{\mathbf{W}}_{a_{j}} \displaybreak[3] \nonumber \\
&\qquad +\frac{\gamma^{-2}}{2}h(\mathbf{x})h^{T}(\mathbf{x})\nabla \boldsymbol{\phi}_{v_{i}}^{T}\tilde{\mathbf{W}}_{a_{i}}+p(\mathbf{x})\tilde{\boldsymbol{\nu}}\Big)\Bigg)  \nonumber \\
&\qquad -\sum_{i=1}^{N} \tilde{\mathbf{W}}_{a_{i}}^{T}\Big[ \theta_{i}^{-1}\dot{\hat{\mathbf{W}}}_{a_{i}}-\frac{1}{4}\sum_{k=1}^{N}C_{i}^{k}\nonumber \\
&\qquad \times \hat{\mathbf{W}}_{a_{i}}\boldsymbol{\mu}_{k}^{T}\hat{\mathbf{W}}_{v_{k}} +\frac{1}{4}E_{i}\hat{\mathbf{W}}_{a_{i}}\boldsymbol{\mu}_{i}^{T}\hat{\mathbf{W}}_{v_{i}} \Big] \nonumber \\
&\qquad  -\tilde{\mathbf{W}}_{a_{N+1}}^{T}\Big[\theta_{N+1}^{-1}\dot{\hat{\mathbf{W}}}_{a_{N+1}}\nonumber \\
&\qquad +\frac{1}{4}E_{N+1}\hat{\mathbf{W}}_{a_{N+1}}\boldsymbol{\mu}_{N+1}^{T}\hat{\mathbf{W}}_{v_{N+1}} \Big] 
\end{align}
where $ \check{\boldsymbol{\nu}} -\hat{\boldsymbol{\nu}}=\tilde{\boldsymbol{\nu}} $. So update rules for action networks are defined as:
\begin{align}
\dot{\hat{\mathbf{W}}}_{a_{i}}&=-\theta_{i}\Big[ (F_{2}^{i}\hat{\mathbf{W}}_{a_{i}}-F_{1}^{i}\bar{\boldsymbol{\eta}}_{i}^{T} \hat{\mathbf{W}}_{v_{i}}) -\frac{1}{4}\sum_{k=1}^{N} C_{i}^{k}\hat{\mathbf{W}}_{a_{i}}\nonumber \\
&\qquad \times \boldsymbol{\mu}_{k}^{T}\hat{\mathbf{W}}_{v_{k}} +\frac{1}{4}E_{i}\hat{\mathbf{W}}_{a_{i}}\boldsymbol{\mu}_{i}^{T}\hat{\mathbf{W}}_{v_{i}} \Big] \quad 1\leq i\leq N
\end{align}
\begin{align}
\dot{\hat{\mathbf{W}}}_{a_{N+1}}&=-\theta_{N+1}\Big[ (F_{2}^{N+1}\hat{\mathbf{W}}_{a_{N+1}}-F_{1}^{N+1}\bar{\boldsymbol{\eta}}_{N+1}^{T} \hat{\mathbf{W}}_{v_{N+1}})\nonumber \\
&\qquad +\frac{1}{4}E_{N+1}\hat{\mathbf{W}}_{a_{N+1}}\boldsymbol{\mu}_{N+1}^{T}\hat{\mathbf{W}}_{v_{N+1}} \Big]
\end{align}
where $ F_{2}^{i}$s and $ F_{1}^{i}$s are tuning parameters and they should be selected as mentioned in proceed of proof. With these tuning laws the following terms added to $ \dot{L}(\mathbf{x})$:
\begin{equation}
\sum_{i=1}^{N+1}\tilde{\mathbf{W}}_{a_{i}}^{T} \Big( F_{2}^{i}\mathbf{W}_{v_{i}}-F_{2}^{i}\tilde{\mathbf{W}}_{a_{i}} -F_{1}^{i}\bar{\boldsymbol{\eta}}_{i}^{T}\mathbf{W}_{v_{i}}+F_{1}^{i}\bar{\boldsymbol{\eta}}_{i}^{T}\tilde{\mathbf{W}}_{v_{i}}\Big)
\end{equation}
Now \eqref{ldot} becomes
\begin{align}
& \dot{L}(\mathbf{x}) = \sum_{i=1}^{N+1}\Bigg( -Q_{i}(\mathbf{x})+\epsilon_{HJ_{i}}-\frac{1}{4}\sum_{j=1}^{N}\mathbf{W}_{v_{j}}^{T}C_{j}^{i}\mathbf{W}_{v_{j}}\nonumber \\
&\qquad +\frac{1}{4}\sum_{j=1}^{N} \tilde{\mathbf{W}}_{a_{j}}^{T}C_{j}^{i}\tilde{\mathbf{W}}_{a_{j}}\boldsymbol{\mu}_{i}^{T}\mathbf{W}_{v_{i}} \nonumber \\
&\qquad +\frac{1}{4}\mathbf{W}_{v_{i}}^{T}E_{i}(\mathbf{W}_{v_{i}}-2\tilde{\mathbf{W}}_{a_{i}}) - \tilde{\mathbf{W}}_{v_{i}}^{T}\bar{\boldsymbol{\eta}}_{i}\bar{\boldsymbol{\eta}}_{i}^{T}\tilde{\mathbf{W}}_{v_{i}}\nonumber \\
&\qquad -\frac{1}{2}\sum_{\substack{j=1 \\ j\neq i}}^{N}\tilde{\mathbf{W}}_{a_{j}}^{T}(C_{j}^{i}-D_{i}^{j})\mathbf{W}_{v_{j}}\boldsymbol{\mu}_{i}^{T}\tilde{\mathbf{W}}_{v_{i}}+\nabla \epsilon_{i}^{T}\mathbf{f}(\mathbf{x}) \nonumber \\
&\qquad -\mathbf{W}_{v_{i}}^{T}\nabla \boldsymbol{\phi}_{v_{i}} p(\mathbf{x})\tilde{W}_{\nu}^{T}\boldsymbol{\phi}_{\nu}+ \boldsymbol{\phi}_{\nu}^{T}\tilde{W}_{\nu}S_{i}(\tilde{W}_{\nu}-2W_{\nu})^{T}\nonumber \\
&\qquad \times \boldsymbol{\phi}_{\nu}\boldsymbol{\mu}_{i}^{T}\tilde{\mathbf{W}}_{v_{i}}-\tilde{\mathbf{W}}_{v_{i}}^{T}\boldsymbol{\mu}_{i}\mathbf{W}_{v_{i}}^{T}\nabla \boldsymbol{\phi}_{v_{i}} p(\mathbf{x})\tilde{W}_{\nu}^{T}\boldsymbol{\phi}_{\nu}
\nonumber \\
&\qquad +\frac{1}{4}\tilde{\mathbf{W}}_{a_{i}}^{T}E_{i}(\mathbf{W}_{v_{i}}-\tilde{\mathbf{W}}_{a_{i}})\boldsymbol{\mu}_{i}^{T}\mathbf{W}_{v_{i}}-\boldsymbol{\phi}_{\nu}^{T}W_{\nu}S_{i}\nonumber \\
&\qquad \times W_{\nu}^{T}\boldsymbol{\phi}_{\nu}+\frac{1}{2}\mathbf{W}_{v_{i}}^{T}\sum_{j=1}^{N}D_{i}^{j}\tilde{\mathbf{W}}_{a_{j}} \nonumber \\
&\qquad +\frac{1}{4}\sum_{j=1}^{N} \tilde{\mathbf{W}}_{a_{j}}^{T}C_{j}^{i}\mathbf{W}_{v_{j}}\boldsymbol{\mu}_{i}^{T}(\tilde{\mathbf{W}}_{v_{i}}-\mathbf{W}_{v_i})\nonumber \\
&\qquad -\frac{1}{4}\tilde{\mathbf{W}}_{a_{i}}^{T}E_{i}\mathbf{W}_{v_{i}}\boldsymbol{\mu}_{i}^{T}\tilde{\mathbf{W}}_{v_{i}}+\tilde{\mathbf{W}}_{v_{i}}^{T}\boldsymbol{\mu}_{i}\epsilon_{HJ_{i}} \nonumber \\
&\qquad +\tilde{\mathbf{W}}_{a_{i}}^{T} \Big( F_{2}^{i}\mathbf{W}_{v_{i}}-F_{2}^{i}\tilde{\mathbf{W}}_{a_{i}}-F_{1}^{i}\bar{\boldsymbol{\eta}}_{i}^{T}\mathbf{W}_{v_{i}} \displaybreak[3] \nonumber \\
&\qquad +F_{1}^{i}\bar{\boldsymbol{\eta}}_{i}^{T}\tilde{\mathbf{W}}_{v_{i}}\Big)+\nabla \epsilon_{i}^{T}\Big(+p(\mathbf{x})(W_{\nu}-\tilde{W}_{\nu})^{T}\boldsymbol{\phi}_{\nu} \nonumber \\
&\qquad -\frac{1}{2}\sum_{j=1}^{N}g_{j}(\mathbf{x})R_{jj}^{-1}g_{j}^{T}(\mathbf{x})\nabla \boldsymbol{\phi}_{v_{j}}^{T}(\mathbf{W}_{v_{j}}-\tilde{\mathbf{W}}_{a_{j}})\nonumber \\
&\qquad +\frac{\gamma^{-2}}{2}h(\mathbf{x})h^{T}(\mathbf{x})\nabla \boldsymbol{\phi}_{v_{i}}^{T}(\mathbf{W}_{v_{i}}-\tilde{\mathbf{W}}_{a_{i}})\Big) \Bigg)
\end{align}

A new parameter is defined which only has ideal weights:
\begin{align}
\Upsilon &=\sum_{i=1}^{N+1}\Bigg( \epsilon_{HJ_{i}}-\frac{1}{4}\sum_{j=1}^{N}\mathbf{W}_{v_{j}}^{T}C_{j}^{i}\mathbf{W}_{v_{j}}+\frac{1}{4}\mathbf{W}_{v_{i}}^{T}E_{i}\mathbf{W}_{v_{i}}\nonumber \\
&\qquad -\boldsymbol{\phi}_{\nu}^{T}W_{\nu}S_{i}W_{\nu}^{T}\boldsymbol{\phi}_{\nu}+\nabla \epsilon_{i}^{T}p(\mathbf{x})W_{\nu}^{T}\boldsymbol{\phi}_{\nu}\displaybreak[3] \nonumber \\
&\qquad +\nabla \epsilon_{i}^{T}\Big(-\frac{1}{2}\sum_{j=1}^{N}g_{j}(\mathbf{x})R_{jj}^{-1}g_{j}^{T}(\mathbf{x})\nabla \boldsymbol{\phi}_{v_{j}}^{T}\mathbf{W}_{v_{j}}\nonumber \\
&\qquad +\frac{\gamma^{-2}}{2}h(\mathbf{x})h^{T}(\mathbf{x})\nabla \boldsymbol{\phi}_{v_{i}}^{T}\mathbf{W}_{v_{i}}\Bigg)
\end{align}
According to the assumptions, $\Upsilon$ has an upper bound, $\Upsilon_{max}$.
Since $ \epsilon_{HJ_{i}}$s are NN approximations errors of value functions, $\kappa$ (number of hidden layers in NNs) can be selected such that to decrease these errors to a desired constant:
\begin{equation*}
\forall \epsilon_{i}>0, \; \exists \kappa^{*}\; s.t.\; \; \; \sup_{x\in \Omega}\Vert \epsilon_{HJ_{i}}\Vert <\epsilon_{i}\; \; \; for\; \; \; \kappa >\kappa^{*}
\end{equation*}
Now by assuming $ \kappa >\kappa^{*}$  and defining $ \tilde{Z}_{1}=\mathbf{x}$, $ \tilde{Z}_{2}=[ \tilde{\mathbf{W}}_{v_1}^{T}\bar{\boldsymbol{\eta}}_{1}, \dots ,  \tilde{\mathbf{W}}_{v_{N+1}}^{T}\bar{\boldsymbol{\eta}}_{N+1}]^{T} $, $ \tilde{Z}_{3}=[ \tilde{\mathbf{W}}_{a_1}^{T}, \dots , \tilde{\mathbf{W}}_{a_{N+1}}^{T}]^{T}$, $ \tilde{Z}_{4}= \tilde{W}_{\nu}^{T}\boldsymbol{\phi}_{\nu} $, and $ \tilde{Z}=[\tilde{Z}_{1}^{T}, \tilde{Z}_{2}^{T}, \tilde{Z}_{3}^{T}, \tilde{Z}_{4}^{T}]^{T}$, \eqref{ldot} can be rewritten as
\begin{align}
\dot{L}(\mathbf{x}) &= \Upsilon +\tilde{Z}_{2}^{T}\bigg(\tilde{Z}_{4}^{T}\otimes I_{N+1}\bigg)S\tilde{Z}_{4} -\tilde{Z}^{T}M\tilde{Z}\nonumber \\
&\qquad +\tilde{Z}^{T}\begin{bmatrix}
\lambda_{1}\\
\lambda_{2}\\
\vdots \\
\lambda_{2N+4}
\end{bmatrix}+\nabla \epsilon_{i}^{T}\mathbf{f}(\mathbf{x})
\end{align}
Where $ \otimes $ is the Kronecker product, $ I_{N+1}$ is the identity matrix and $ S=[S_{1}^{T}, S_{2}^{T}, \dots , S_{N+1}^{T}]^{T}$. By assuming $ 2\leq a\leq N+2 $, $ N+3\leq b\leq 2N+2 $, $ i=a-1 $ and $ j=b-N-2 $, the blocks of symmetric matrix $M$ and vector $ \Lambda $ are:
\begin{align}
&m_{11}=\sum_{l=1}^{N+1}\vartheta_{l} \nonumber \\
&m_{aa}=1 \nonumber \\
&m_{ba}=\frac{1}{8\rho_i}C_{j}^{i}\mathbf{W}_{v_{j}}-\frac{1}{4\rho_i}D_{i}^{j}\mathbf{W}_{v_i}, \qquad i\neq j \nonumber \\
&m_{ba}=-\frac{1}{8\rho_j}D_{j}^{j}\mathbf{W}_{v_j}+\frac{1}{8}E_{j}\mathbf{W}_{v_j}-\frac{1}{2}F_{1}^{j}, \qquad i=j \nonumber \\
& m_{bb}=-\frac{1}{4}\sum_{l=1}^{N+1}C_{j}^{l}\boldsymbol{\mu}_{l}^{T}\mathbf{W}_{v_{l}}\nonumber \\
&\qquad +\frac{1}{4}E_{j}\boldsymbol{\mu}_{j}^{T}\mathbf{W}_{v_{j}}+F_{2}^{j} \nonumber \\
&m_{a(2N+4)}=\frac{1}{2\rho_i} \mathbf{W}_{v_{i}}^{T}\nabla \boldsymbol{\phi}_{v_{i}}p(\mathbf{x})+\frac{1}{\rho_i}\boldsymbol{\phi}_{\nu}^{T}\mathbf{W}_{\nu}S_{i} \nonumber \\
&m_{(2N+3)(2N+3)}=\frac{1}{4}E_{N+1}\boldsymbol{\mu}_{N+1}^{T}+F_{2}^{N+1} \displaybreak[3]\nonumber \\
& m_{1a}=m_{1b}=m_{1(2N+3)}=m_{1(2N+4)}=m_{b(2N+3)} \nonumber \\
& \quad \quad =m_{b(2N+4)}=m_{(2N+3)(2N+4)}\nonumber \\
& \quad \quad =m_{(2N+4)(2N+4)}=\mathbf{0}
\end{align}
\begin{equation*}
m_{a(2N+3)}= \left\{ \begin{array}{lr}
\frac{1}{8}E_{N+1}\mathbf{W}_{v_{N+1}}-\frac{1}{2}F_{1}^{N+1} , & a=N+2 \\
\mathbf{0} , & a\neq N+2
\end{array} \right.
\end{equation*}
and
\begin{align}
\lambda_{1}&=\mathbf{0} \label{d1} \\
\lambda_{a}&=\frac{\epsilon_{HJ_{i}} }{\rho_i} \nonumber \\
\lambda_{b}&=\frac{1}{2}\sum_{l=1}^{N+1}\big(D_{l}^{j}\big)^{T}\mathbf{W}_{v_l}-\frac{1}{2}E_{j}\mathbf{W}_{v_j}\nonumber \\
&\; -\frac{1}{4}\sum_{l=1}^{N+1}C_{j}^{l}\boldsymbol{\mu}_{l}^{T}\mathbf{W}_{v_{l}}+F_{2}^{j}\mathbf{W}_{v_j} \nonumber \\
&\; +\sum_{l=1}^{N+1}\frac{1}{2}\nabla \boldsymbol{\phi}_{v_{j}}g_{j}(\mathbf{x})R_{jj}^{-1}R_{lj}R_{jj}^{-1}\nonumber \\
&\; \times g_{j}^{T}(\mathbf{x})\nabla \epsilon_{l} -\frac{\gamma^{-2}}{2}\nabla \boldsymbol{\phi}_{v_{j}}h(\mathbf{x})h^{T}(\mathbf{x})\nabla \epsilon_{j} \nonumber \\
&\; +\frac{1}{4}E_{j}\mathbf{W}_{v_j}\boldsymbol{\mu}_{j}^{T}\mathbf{W}_{v_{j}}-F_{1}^{j}\bar{\boldsymbol{\eta}}_{j}^{T}\mathbf{W}_{v_j} \nonumber \\
\lambda_{2N+3}&=-\frac{1}{2}E_{N+1}\mathbf{W}_{v_{N+1}}+F_{2}^{N+1}\mathbf{W}_{v_{N+1}} \nonumber \\
&\; +\frac{1}{4}E_{N+1}\mathbf{W}_{v_{N+1}}\boldsymbol{\mu}_{N+1}^{T}\mathbf{W}_{v_{N+1}} \displaybreak[3] \nonumber \\
&\; -\frac{\gamma^{-2}}{2}\nabla \boldsymbol{\phi}_{v_{N+1}}h(\mathbf{x})h^{T}(\mathbf{x})\nabla \epsilon_{N+1} \nonumber \\
&\; -F_{1}^{N+1}\bar{\boldsymbol{\eta}}_{N+1}^{T}\mathbf{W}_{v_{N+1}}\nonumber \\
\lambda_{2N+4}&=-\sum_{l=1}^{N+1}p^{T}(\mathbf{x})\nabla \boldsymbol{\phi}^{T}_{v_l}\mathbf{W}_{v_l}-p^{T}(\mathbf{x})\sum_{l=1}^{N+1}\nabla \epsilon_{l}
\end{align}
Since $f(\mathbf{x})$ is Lipschitz, one can get
\begin{align}
\dot{L}(\mathbf{x}) &\leq \Upsilon +\tilde{Z}_{2}^{T}\bigg(\tilde{Z}_{4}^{T}\otimes I_{N+1}\bigg)S\tilde{Z}_{4} -\tilde{Z}^{T} M\tilde{Z} \nonumber \\
&\qquad +\tilde{Z}^{T}\begin{bmatrix}
\lambda_{1}\\
\lambda_{2}\\
\vdots \\
\lambda_{2N+4}
\end{bmatrix}
\end{align}
and \eqref{d1} becomes $ \lambda_{1}=\sum_{l=1}^{N+1}b_{\epsilon_{x_l}}b_{f} $.
Parameters $ F_{2}^{i} $ and $ F_{1}^{i} $ should be chosen such that matrix $ M$ becomes positive definite. Elements of vector $\Lambda$ are bounded. Thus:
\begin{equation}
\dot{L}<-\Vert \tilde{Z}\Vert^{2}\lambda_{min}(M)+\Lambda_{max}\Vert \tilde{Z}\Vert +\Upsilon_{max}+\Vert \tilde{Z}_{2}\Vert \Vert \tilde{Z}_{4}^{2}\Vert \Vert S\Vert \label{ineq}
\end{equation}
Expanding terms in \eqref{ineq} yields:
\begin{align}
&\dot{L}<-\lambda_{min}(M)\big(\Vert \tilde{Z}_1\Vert^{2}+\Vert \tilde{Z}_2\Vert^{2}+\Vert \tilde{Z}_3\Vert^{2}+\Vert \tilde{Z}_4\Vert^{2}\big) \nonumber \\
&\quad +\Lambda_{max}\sqrt{\Vert \tilde{Z}_1\Vert^{2}+\Vert \tilde{Z}_2\Vert^{2}+\Vert \tilde{Z}_3\Vert^{2}+\Vert \tilde{Z}_4\Vert^{2}}\displaybreak[3] \nonumber \\
&\quad +\Vert \tilde{Z}_{2}\Vert \Vert \tilde{Z}_{4}^{2}\Vert \Vert S\Vert +\Upsilon_{max} \nonumber \\
&\; < -\lambda_{min}(M)\big(\Vert \tilde{Z}_1\Vert^{2}+\Vert \tilde{Z}_2\Vert^{2}+\Vert \tilde{Z}_3\Vert^{2}+\Vert \tilde{Z}_4\Vert^{2}\big) \nonumber \\
&\quad +\Lambda_{max}\big(\Vert \tilde{Z}_1\Vert +\Vert \tilde{Z}_2\Vert +\Vert \tilde{Z}_3\Vert +\Vert \tilde{Z}_4\Vert \big) \nonumber \\
&\quad +\Vert \tilde{Z}_{2}\Vert \Vert \tilde{Z}_{4}\Vert^{2} \Vert S\Vert +\Upsilon_{max}
\end{align}
For simplicity $ \lambda_{min}(M)=\lambda_{m} $, $ \Lambda_{max}=\Lambda_{m} $, $ \Upsilon_{max}=\Upsilon_{m} $ and $ \Vert S\Vert =S_{m} $. By defining $ \Vert \tilde{Z}_{2}\Vert =r\sin(\varphi) $ and $ \Vert \tilde{Z}_{4}\Vert =r\cos(\varphi) $ where $ 0\leq \varphi \leq \frac{\pi}{2} $ , one can get:
\begin{align}
& \qquad \Lambda_{m}\big(\Vert \tilde{Z}_2\Vert +\Vert \tilde{Z}_4\Vert \big)+\Vert \tilde{Z}_2\Vert \Vert \tilde{Z}_4\Vert^{2}S_{m}\nonumber \\
& \qquad +\lambda_{m}(M)\big(\Vert \tilde{Z}_2\Vert^{2}+\Vert \tilde{Z}_4\Vert^{2}\big)
&= \label{124} \\
& \qquad -\lambda_{m}r^{2}+\Lambda_{m} r\big( \sin(\varphi)+\cos(\varphi)\big) \nonumber \\
&\qquad +S_{m}r^{3}\sin(\varphi)\cos^{2}(\varphi) 
& \leq \nonumber \\
&\qquad -\lambda_{m}r^{2}+\sqrt{2}r\Lambda_{m} + \big( 0.385 S_{m}\big)r^{3}
\end{align}
Thus for
\begin{equation}\label{111}
r_1 < r < r_2
\end{equation}
where
\begin{align}
r_1=& \frac{\lambda_{m} -\sqrt{\lambda_{m}^{2}-1.54\sqrt{2}S_{m}\Lambda_{m}}}{0.77 S_{m}}\nonumber \\
r_2=& \frac{\lambda_{m} +\sqrt{\lambda_{m}^{2}-1.54\sqrt{2}S_{m}\Lambda_{m}}}{0.77 S_{m}} \nonumber 
\end{align}
 relation \eqref{124} becomes negative. Tuning parameters $ F_{2}^{i} $s and $ F_{1}^{i} $s should be selected such that $ \lambda_{m} $ becomes large enough to achieve two goals: making the term under radical sign positive and making the region in \eqref{111} big enough to discard improper initializations of $ \hat{\mathbf{W}}_{v_i} $s and $ \hat{W}_{\nu} $.
 Now if
\begin{equation}
\Vert \tilde{Z}_1\Vert +\Vert \tilde{Z}_3\Vert >\frac{\Lambda_{m}}{\lambda_{m}}+\sqrt{\frac{\Upsilon_{m}}{\lambda_{m}}+\frac{\Lambda_{m}^{2}}{2\lambda_{m}^2}}
\end{equation}
with condition \eqref{111}, $ \dot{L} $ becomes negative. So with proper initialization, $ \mathbf{x} $, $ \hat{\mathbf{W}}_{v_i} $s and $ \hat{\mathbf{W}}_{a_i} $s and hence $ \hat{\mathbf{W}}_{v_i}-\hat{\mathbf{W}}_{a_i} $s become UUB. According to \eqref{nuhati} and using gradient descent for updating $ \hat{W}_{\nu} $ and ultimate boundedness of $ \hat{\mathbf{W}}_{a_i} $s and $ \mathbf{x} $, $ \hat{W}_{\nu} $ becomes UUB too.
\end{IEEEproof}



%
%

\ifCLASSOPTIONcaptionsoff
  \newpage
\fi

\bibliographystyle{ieeetr}
\bibliography{MyReferences}
\end{document}